\documentclass[12pt]{article}

\usepackage{newtxtext,newtxmath}
\usepackage{graphicx}
\usepackage[letterpaper,margin=1in]{geometry}
\renewenvironment{abstract}{\quotation}{\endquotation}
\date{}
\renewcommand\refname{References and Notes}

\makeatletter
\renewcommand{\fnum@figure}{\textbf{Figure \thefigure}}
\renewcommand{\fnum@table}{\textbf{Table \thetable}}
\makeatother

\usepackage{scicite}
\usepackage{url}

\def\scititle{Oxygen stoichiometry directs rutile--anatase phase selection through kinetic control of nucleation}
\title{\bfseries \boldmath \scititle}

\author{
  Han Uk~Lee$^{1}$,
  Hyeon Woo~Kim$^{2}$,
  Ji Min~Kim$^{1}$,\and
  Dong Won~Jeon$^{1}$,
  Rohan~Mishra$^{2\ast}$,
  Sung Beom~Cho$^{1\ast}$\and
  \small$^{1}$ School of Advanced Materials Science and Engineering, Sungkyunkwan University, Suwon, Republic of Korea\and
  \small$^{2}$ Department of Mechanical Engineering \& Materials Science, Washington University in St. Louis, USA\and
  \small$^{\ast}$Corresponding author. Email: rmishra@wustl.edu; sungcho@skku.edu
}

\newcommand{\sciauthorline}{Han Uk~Lee, Hyeon Woo~Kim, Ji Min~Kim, Dong Won~Jeon, Rohan~Mishra$^{\ast}$, Sung Beom~Cho$^{\ast}$}

\begin{document}

\maketitle

\begin{abstract} \bfseries \boldmath
Synthesis of a target polymorph remains more empirical than predictive because crystallization often selects the most accessible nucleation pathway rather than the thermodynamically most stable phase. Here, we show that oxygen stoichiometry converts this empirical synthesis variable into a kinetic control parameter for anatase--rutile selection in TiO$_{2-x}$. Enhanced-sampling simulations reveal that oxygen content alters the nucleation-barrier landscape, switching the relative accessibility of anatase and rutile, even while rutile remains thermodynamically favored. Molecular dynamics simulations show the presence of a diffuse intermediate shell around the nucleus, where oxygen deficiency alters Ti--O coordination and connectivity and drives shell-local motif evolution from anatase-like toward rutile-like environments. A coupled-flux model that integrates barrier competition with shell-mediated attachment/exchange yields a relative nucleation-rate map consistent with reported oxygen-dependent synthesis trends. These results establish stoichiometry-controlled intermediate-shell motif evolution as a kinetic origin of polymorph selection and provide a framework for predicting target phases in composition-coupled crystallization.
\end{abstract}

\noindent
Controlling a target polymorph is a central challenge in materials science because the selected phase governs functional performance, long-term stability, and process compatibility in practical applications. Despite the long-standing recognition of crystal polymorphism, predicting which polymorph will form under a given synthesis condition remains difficult. Equilibrium thermodynamic descriptions, including formation energies, phase diagrams, and convex-hull analyses, provide a useful baseline for anticipating stable phases, but thermodynamic stability alone does not guarantee experimental accessibility~\cite{sun2016thermodynamic}. Indeed, in diverse systems, including TiO$_2$~\cite{hanaor2011review}, Ga$_2$O$_3$~\cite{bosi2020ga,gann2022initial}, and CaCO$_3$~\cite{arvidson1999dolomite,kim2023dissolution}, the polymorph obtained experimentally can differ from the thermodynamic stability ranking. This gap arises because crystallization is initiated by nucleation events that are themselves governed by transient nanoscale structural and compositional fluctuations, rather than by bulk equilibrium stability alone~\cite{kelton2010nucleation}. As a result, practical polymorph control often still relies on empirical synthesis windows, chemical intuition, and system-specific heuristics rather than predictive theory~\cite{kocevski2023predicting}. These observations indicate that polymorph selection should not be viewed solely as an equilibrium ranking among competing phases, but as a kinetic competition among nucleation pathways. A predictive description of target-polymorph selection therefore needs to connect equilibrium stability with pathway accessibility under synthesis conditions.

Classical nucleation theory (CNT) has long provided the canonical framework for describing nucleation kinetics~\cite{becker1935kinetische}. In its standard form, CNT treats nucleation as the emergence of a new-phase nucleus from an effectively homogeneous parent phase, with the nucleus separated from the parent phase by a sharp interface~\cite{debenedetti2024special}. This picture has been successful as a baseline for many crystallization problems~\cite{miller1983homogeneous,strey1994problem,manka2010homogeneous}, but it becomes restrictive when the nucleus--parent region is spatially diffuse or structurally heterogeneous~\cite{sosso2016crystal,granasy2002diffuse}. In many systems, atoms near a nucleus do not change abruptly from parent-like to crystal-like environments. Instead, the surrounding region can form a diffuse intermediate shell in which the local composition, coordination, and other structural order parameters evolve continuously during nucleation, as illustrated in Fig. 1~\cite{li2020revealing,hu2020physical,kawasaki2010formation,tan2014visualizing,peng2015two,zhou2019observing,lin2023situ}. Such a shell is not merely a geometric transition zone: because atoms attach to, detach from, exchange with, and rearrange around the nucleus through this region, its local state can influence the kinetic behavior of the nucleation pathway. Related precursor states, amorphous or partially ordered intermediates, and diffuse transition regions have been reported in diverse systems, including colloids~\cite{peng2015two}, FePt~\cite{zhou2019observing}, metallic glasses~\cite{sheng2026classical}, and YAG~\cite{lin2023situ}. For polymorph selection, these intermediate regions are especially important because their local structural and compositional states can bias both pathway accessibility and the kinetic processes by which a nucleus grows or dissolves. A predictive description of polymorphic competition should therefore account not only for the free-energy barrier of each pathway, but also for shell-mediated kinetic processes that modulate the relative nucleation rates of competing polymorphs.

To address this gap, we present a nucleation-based framework showing how oxygen stoichiometry modulates both nucleation barriers and shell-mediated kinetic biases, thereby controlling the relative nucleation rates of competing phases. We use anatase--rutile selection in TiO$_{2-x}$ as a representative case because oxygen stoichiometry is an empirical synthesis handle known to bias TiO$_2$ polymorph outcomes~\cite{mangum2020utilizing,palmolahti2024production,li2016understanding,rafieian2015controlled,saari2022low}. Using our framework, we show that anatase--rutile phase selection emerges from the relative accessibility of competing nucleation pathways under synthesis conditions, rather than from thermodynamic stability alone. Specifically, we first translate oxygen stoichiometry into a nucleation-barrier landscape for anatase--rutile competition. We then identify a diffuse nucleus--parent transition shell and evaluate whether its local coordination and polymorph-like structural motifs bias anatase or rutile pathways. Finally, we incorporate these barrier-level and local shell contributions into a coupled-flux model to obtain a relative nucleation-rate map for the two polymorphs. By comparing this map with experimentally reported phase-selection trends, this work provides a nucleation-theory-based route for connecting empirical synthesis control to target-polymorph prediction.

\section*{Results and Discussion}
Anatase--rutile competition in TiO$_2$ has been extensively reported experimentally, but remains poorly understood. Thus, it provides a useful model for examining polymorph selection. The two phases share TiO$_6$ octahedral building blocks but differ in network connectivity and thermodynamic stability. In anatase, each TiO$_6$ octahedron shares four edges and four corners with neighboring octahedra, whereas in rutile each octahedron shares two edges and eight corners (Fig. 2A). This difference in octahedral connectivity distinguishes the two polymorphs structurally and provides the motif-level basis for analyzing pathway preference during nucleation. The two phases also exhibit distinct functional characteristics, making polymorph selection important for applications ranging from photocatalysis to optical and high-$k$ dielectric materials~\cite{fujishima2000titanium,robertson2015high,jeon2026rutile}. From the perspective of thermodynamic equilibrium, the expected phase preference is straightforward. As shown in Fig. 2B, rutile remains lower in Gibbs free energy than anatase over the temperature range examined, consistent with the conventional understanding that rutile is the thermodynamically stable phase of TiO$_2$, whereas anatase is metastable~\cite{smith2009heat,banfield1998thermodynamic,muscat2002first}. If equilibrium thermodynamics alone controlled polymorph selection, rutile would be expected to dominate throughout the temperature range examined. The key question is whether this expectation holds under practical synthesis conditions.

To examine whether this equilibrium preference is reflected experimentally, we compiled literature-reported anatase--rutile formation outcomes in temperature--pressure space (Fig. 2C). Although rutile is thermodynamically more stable over the examined range, anatase is frequently obtained at lower synthesis temperatures. Rutile becomes more prevalent at higher temperatures, while elevated pressure shifts the reported anatase--rutile transition boundary to lower temperatures. The reported transition temperatures span a broad range, approximately 400--1200~$^\circ$C, depending on the synthesis method and experimental protocol~\cite{karunadasa2022microstructural,byrne2016new,vahldiek1966phase}. This broad and condition-dependent variation shows that the experimentally selected polymorph does not simply follow the equilibrium Gibbs free-energy ranking.

Among the synthesis variables that can shift anatase--rutile phase selection, oxygen stoichiometry provides a particularly useful control axis. Oxygen-deficient TiO$_{2-x}$ has been empirically associated with enhanced rutile formation, whereas oxygen-rich or oxidizing conditions tend to favor anatase formation~\cite{mangum2020utilizing,palmolahti2024production,li2016understanding,rafieian2015controlled,saari2022low}. To test whether this stoichiometry-dependent trend can be explained by equilibrium phase stability, we calculated the grand potential of anatase and rutile. Across the accessible range of oxygen conditions, it favors rutile or reduced Ti--O phases (e.g., Magnéli phases) over anatase (see Supplementary Note 1). Therefore, the observed stoichiometry-dependent polymorph preference is unlikely to arise from equilibrium phase stability. This motivates a kinetic framework that connects oxygen stoichiometry to anatase--rutile nucleation-pathway competition.

To quantify this stoichiometry-dependent competition, we express anatase--rutile selection in TiO$_{2-x}$ in terms of the relative nucleation rates of the two pathways. For a polymorph $i$ nucleating from the disordered TiO$_{2-x}$ parent phase, the nucleation rate can be written as
\begin{equation}
J_i = A_i \exp\left(-\frac{\Delta G_i^*}{k_\mathrm{B}T}\right),
\qquad i \in {\mathrm{anatase}, \mathrm{rutile}}.
\label{eq:nucleation-rate}
\end{equation}
\noindent
Here, $\Delta G_i^*$ is the reversible free-energy cost required to form a critical nucleus of polymorph $i$, and therefore determines the barrier that must be overcome along that pathway. The prefactor $A_i$ collects the kinetic factors associated with that pathway, including atomic attachment, detachment, and local structural rearrangement near the nucleus. Thus, understanding stoichiometry-dependent polymorph selection requires determining how oxygen composition modifies both the nucleation barrier, $\Delta G_i^*$, and the pathway-dependent kinetic contribution represented by $A_i$. We therefore first evaluate the stoichiometry dependence of $\Delta G_i^*$ using enhanced-sampling simulations, and then examine how oxygen composition alters the near-nucleus environments that can contribute to $A_i$ through the analysis of molecular dynamics (MD) simulations. The resulting relative nucleation rates provide a kinetic measure of anatase--rutile polymorph preference. Details of the simulation protocols are provided in the Materials and Methods and Supplementary Text.

We first quantify how oxygen stoichiometry affects the nucleation-barrier landscape for the anatase--rutile competition. In the classical picture, this barrier reflects the balance between the chemical-potential driving force for forming the new phase and the interfacial penalty associated with creating a nucleus--parent boundary. Accordingly, the barrier increases with the effective nucleus--parent interfacial energy, $\gamma_i$, and decreases with the parent-to-polymorph chemical-potential difference, $|\Delta\mu_{\ell\to i}|$. This scaling can be written as
\begin{equation}
\Delta G_i^*
\propto
\frac{\gamma_i^3}{\left|\Delta\mu_{\ell\to i}\right|^2},
\qquad
i\in{\mathrm{anatase},\mathrm{rutile}}.
\label{eq:barrier-scaling}
\end{equation}
\noindent
Equivalently, at the critical cluster size $n_{c,i}$, the barrier can be approximated as
\begin{equation}
\Delta G_i^*
\simeq
\frac{1}{2}\left|\Delta\mu_{\ell\to i}\right| n_{c,i},
\qquad
i\in{\mathrm{anatase},\mathrm{rutile}}.
\label{eq:critical-size-barrier}
\end{equation}
\noindent
Here, $\gamma_i$ is the effective nucleus--parent interfacial energy, $|\Delta\mu_{\ell\to i}|$ is the parent-to-polymorph chemical-potential difference, and $n_{c,i}$ is the critical cluster size. In practice, we obtain $|\Delta\mu_{\ell\to i}|$ from enhanced sampling~\cite{branduardi2012metadynamics} and determine $n_{c,i}$ from seeding simulations~\cite{espinosa2016seeding}, which identify the nucleus size at which growth and dissolution are balanced (see Materials and Methods and Supplementary Note 2).

We evaluate the competition between the nucleation of the two phases using the barrier difference, $\Delta G^*_{\mathrm{rutile}}-\Delta G^*_{\mathrm{anatase}}$, plotted as a function of temperature for different oxygen stoichiometries in Fig. 3A. Positive values indicate an anatase-favored barrier ordering, whereas negative values indicate a rutile-favored barrier ordering. Although temperature modulates the barrier difference within each stoichiometry, the transition from anatase-favored to rutile-favored barrier ordering is driven predominantly by the oxygen stoichiometry. These results indicate that oxygen stoichiometry modifies anatase--rutile competition by reordering the barrier-level accessibility of the two nucleation pathways.

We rationalize the observed competition between the nucleation barriers of the two phases by examining the stoichiometry-driven change in local Ti--O polyhedral connectivity, as schematically illustrated in Fig. 3B. Decreasing oxygen stoichiometry introduces oxygen vacancies and lowers the local Ti coordination environment, allowing edge-sharing Ti--O polyhedral motifs to reorganize into more corner-sharing-compatible arrangements. This connectivity change is important for anatase--rutile competition because rutile contains a larger fraction of corner-sharing TiO$_6$ linkages, whereas anatase contains a larger edge-sharing component. Thus, oxygen-deficient environments can make rutile-like connectivity more accessible at the local structural level. The stoichiometry-dependent shift in the nucleation barrier can therefore be interpreted as reflecting changes not only in the local coordination but also in the connectivity motifs that can propagate into an extended polymorph-specific network.

To determine where such connectivity bias emerges during nucleation, we next examine the nucleus--parent transition region. This region is kinetically important because a nucleus near the critical size exchanges atoms with the surrounding parent phase through local attachment, detachment, and structural rearrangements. The nucleus--parent boundary need not be a sharply defined interface; instead, it may form a diffuse intermediate shell in which local composition, coordination, and structural order vary gradually between the crystalline nucleus and the parent phase~\cite{menon2020role,lechner2011role,granasy2002diffuse}. Because nucleation depends not only on the free-energy barrier but also on atomistic processes occurring in this transition region, its structural organization must be characterized to connect local shell evolution to nucleation kinetics. Therefore, we analyze the nucleus--parent transition region in TiO$_{2-x}$ using the Ti coordination number (CN) and environment similarity (ES)~\cite{piaggi2019calculation}, with the goal of identifying how this region contributes to the preference of the anatase--rutile pathway.

We examine the radial profiles of ES and CN from the nucleus center toward the simulation cell boundary for a representative anatase-seeded simulation at 1500~K (Fig. 4A). CN captures a local structural feature, whereas ES measures local crystallinity by comparing the spatial arrangement around each Ti with a polymorph-specific reference environment. The gradual radial decrease in both metrics indicates that the nucleus--parent boundary in TiO$_{2-x}$ is structurally diffuse rather than the sharp interface assumed in standard CNT. We define the intermediate shell as the yellow-shaded region extending from the end of the seed region, approximately 6.5~\AA, to the point at which ES falls below 0.3, approximately 11~\AA. Although the shell boundary is necessarily approximate for such a broadly distributed transition region, its width of approximately 5~\AA{} is similar in scale to sub-nanometer transition regions reported in other systems.~\cite{zhou2019observing,lin2023situ}. Fig. 4B shows a redistribution of coordination environments as oxygen stoichiometry decreases, quantified by relative changes of approximately $-10.8\%$ for CN6 and $+7.7\%$ for CN5 between TiO$_2$ and TiO$_{1.94}$. This coordination shift indicates that oxygen stoichiometry alters the local structure of the intermediate shell. To relate this structural change to polymorph-specific motifs, we next use ES to assess whether atoms within the shell exhibit anatase- or rutile-like local structural character.

This intermediate shell appears to bias polymorph selection at the atomic level (Fig. 4C). Specifically, we first identify crystal-like atoms within the intermediate shell using an ES threshold of 0.5. These atoms are then classified as anatase-like or rutile-like according to whether their local environments more closely match the anatase or rutile reference structure. As oxygen stoichiometry decreases, the anatase-like fraction is progressively suppressed while the rutile-like fraction increases; this trend holds across the temperature range examined, except at 1850~K, where anatase-like ordering is no longer sustained. These trends show that, during nucleation, the intermediate shell is not merely a geometric transition zone but a structurally active region that steers local environments toward polymorph-characteristic motifs. Although this shell-local bias does not by itself determine the final macroscopic phase, it suggests that the intermediate shell can contribute to a kinetic preference toward a particular polymorphic pathway.

Returning to the rate expression introduced above, $J_i=A_i\exp(-\Delta G_i^*/k_\mathrm{B}T)$, the shell analysis indicates that the prefactor $A_i$ cannot be represented solely by a single sharp-interface attachment process. In the standard CNT picture, the nucleus is treated as a new-phase cluster separated from the parent phase by a sharp boundary, and the kinetic contribution is commonly reduced to an attachment rate evaluated at the critical nucleus~\cite{auer2001prediction,niu2020abinitio}. This approximation is useful when nucleation proceeds through a single interface-controlled process. However, the diffuse nucleus--parent transition region observed in TiO$_{2-x}$ contains shell states whose coordination and polymorph-like character vary with oxygen stoichiometry. The kinetic contribution should therefore depend not only on the critical cluster size, but also on the local shell state through which attachment, detachment, exchange, and structural rearrangements occur.

The essential distinction lies in how this nucleus--parent transition region is treated. CNT collapses the microscopic transition region into an effective interface: the nucleation state is described primarily by the cluster size $n$, while the parent phase enters through an interfacial penalty and the kinetic contribution is summarized by the CNT prefactor $A_i^{\mathrm{CNT}}$. However, because this transition region includes an intermediate shell with its own dynamics, the nucleation state is described by both the cluster size $n$ and the shell state $\rho$~\cite{kelton2010nucleation}. Here, $n$ denotes the size of the crystalline cluster, and $\rho$ denotes the local structural and compositional state of the shell. In this description, atoms or structural units do not simply attach across a single ideal interface; instead, exchange with the crystalline cluster is mediated by the intermediate shell.

We account for this effect using a coupled-flux model (CFM) that separates attachment and detachment between the cluster and shell, $k_{i,\pm}(n,\rho)$, from exchange between the shell and the parent phase, $\alpha_i(n,\rho)$ and $\beta_i(n,\rho)$~\cite{kelton2000time}. The $(n,\rho)$ state definitions, master equations, and steady-state derivation are provided in Supplementary Note 3. For the main text, we use its steady-state form evaluated at the critical nucleus, in which the shell-mediated contribution enters through the kinetic prefactor:
\begin{equation}
A_i^{\mathrm{CFM}}
\simeq
\frac{\alpha_i(n_{c,i},0)}
{k_i^+(n_{c,i},1)}
A_i^{\mathrm{CNT}}.
\label{eq:coupled-flux-prefactor}
\end{equation}
Here, $\alpha_i(n_{c,i},0)$ describes the arrival of an atom from the parent phase into an empty shell, and $k_i^+(n_{c,i},1)$ describes the subsequent incorporation of that atom from the shell into the cluster. Thus, the CFM extends the classical prefactor by adding the kinetic effect of shell-mediated exchange.

This formulation helps to clarify how the intermediate shell modifies polymorph competition. In the coupled-flux model, the kinetic prefactor is no longer determined by a single sharp-interface contribution. Instead, it is controlled by the shell-mediated exchange that supplies atoms to the shell and incorporates them into the cluster. Therefore, oxygen stoichiometry can affect the prefactor by changing the local shell environments and the exchange processes associated with those environments. In this sense, the intermediate shell provides a kinetic channel through which local structural evolution can alter polymorph preference beyond the barrier term alone.

We combine the barrier and shell-resolved kinetic contributions to evaluate the relative nucleation preference,
\begin{equation}
\ln\frac{J_{\mathrm{An}}}{J_{\mathrm{Ru}}}
=
-\frac{\Delta G_{\mathrm{An}}^*-\Delta G_{\mathrm{Ru}}^*}{k_\mathrm{B}T}
+
\ln\frac{A_{\mathrm{An}}^{\mathrm{CFM}}}{A_{\mathrm{Ru}}^{\mathrm{CFM}}},
\label{eq:relative-cf-rate}
\end{equation}
where common normalization factors are omitted from the prefactor ratio. Positive values indicate anatase-favored nucleation, whereas negative values indicate rutile-favored nucleation. The first term represents the barrier contribution, while the second term represents the CFM kinetic-prefactor contribution. Thus, the relative-rate expression links the stoichiometry-dependent nucleation barriers in Fig. 3 with the shell-local structural bias identified in Fig. 4.

Using this coupled-flux expression, we quantify anatase--rutile competition in TiO$_{2-x}$ and compare the calculated preference with representative experimental observations (Fig. 5). The relative-rate map, shown on the right y-axis, indicates that oxygen stoichiometry primarily sets the direction of the nucleation preference, while temperature mainly tunes its magnitude. As oxygen stoichiometry decreases from near-stoichiometric TiO$_2$ toward TiO$_{1.94}$, the preferred pathway shifts from anatase to rutile. This trend is consistent with the shell analysis in Fig. 4, where oxygen deficiency suppresses anatase-like shell environments and enhances rutile-like shell environments around the nucleus.

Experimental observations are plotted on the left y-axis of Fig. 5 against oxygen partial pressure. These observations serve as a qualitative synthesis reference rather than as a calibrated composition axis, because $p_{\mathrm{O_2}}$ and oxygen stoichiometry are not one-to-one variables. Nevertheless, $p_{\mathrm{O_2}}$ remains the practical experimental handle used to tune the sample's oxygen content during synthesis. The relative-rate map captures the direction of this reported trend: oxygen-rich conditions align with anatase-favored accessibility, whereas oxygen-poor conditions shift it toward rutile. However, this relative-rate map describes pathway accessibility at the nucleation stage and should not be interpreted as a direct prediction of the final crystallized product. Because the final phase distribution also depends on subsequent growth, we propagate the nucleation preference into growth simulations on composition fields.

Composition fields in real systems may be locally nonuniform~\cite{muller2019relation,tong2023emerging}. We vary the heterogeneity scale of the oxygen-composition field and examine how different composition ($c$)-field structures shape polymorph growth, as shown in Fig. 6. Seed identities are assigned with weights set by the composition-dependent nucleation preference shown in Fig. 5, after which the phase-field model evolves the domains independently (see Supplementary Note 4 for details). Fig. 6 shows the TiO$_2$ case, where finer composition fields produce mostly anatase seed identities and therefore yield anatase-dominated growth, whereas coarser fields allow rutile to emerge locally. Across the remaining stoichiometries, the simulated phase distribution shifts toward rutile as the mean composition becomes more oxygen deficient (Supplementary Figs. S7--S9). At TiO$_{1.94}$, the mean composition favors rutile nucleation: lower-heterogeneity fields accordingly yield rutile-dominated outcomes, whereas higher-heterogeneity fields allow anatase to emerge locally, producing a mixed outcome. These results suggest that microscopic environments can bias the accessibility of competing polymorph pathways and thereby shape the macroscopic phase distribution observed after crystallization, potentially yielding mixed-phase outcomes. Our nucleation-based framework therefore provides a route for linking practical synthesis handles to the nucleation-kinetic landscape that governs polymorph selection.

\section*{Conclusion}
This study frames polymorph control in terms of nucleation theory, where selection is determined by nucleation-pathway competition. Using anatase--rutile competition in TiO$_{2-x}$ as a representative case, we show that oxygen stoichiometry, an empirical synthesis handle for TiO$_2$ polymorph selection, can be recast in terms of the kinetic accessibility of competing nucleation pathways. Within this framework, we found that oxygen stoichiometry alters the kinetic landscape by reordering the nucleation barriers of competing polymorphic pathways. We further identified a diffuse intermediate shell at the nucleus--parent transition region, demonstrating that this region is a structurally active domain. Specifically, oxygen stoichiometry changes the evolution of this shell, biasing its local structure toward polymorph-characteristic motifs and thereby contributing to polymorph preference. Using the coupled-flux model, we integrated these barrier and shell contributions into a relative nucleation-rate map and showed that the resulting framework captures the experimental trends in phase selection. We further extended the nucleation-based framework to growth in heterogeneous composition fields, showing that microscopic environments can influence the final polymorph distribution. By translating polymorph selection into the language of crystallization-pathway accessibility, this work provides a mechanistic basis for target polymorph design under synthesis conditions.


\begin{figure}
  \centering
  \includegraphics[width=1.0\textwidth]{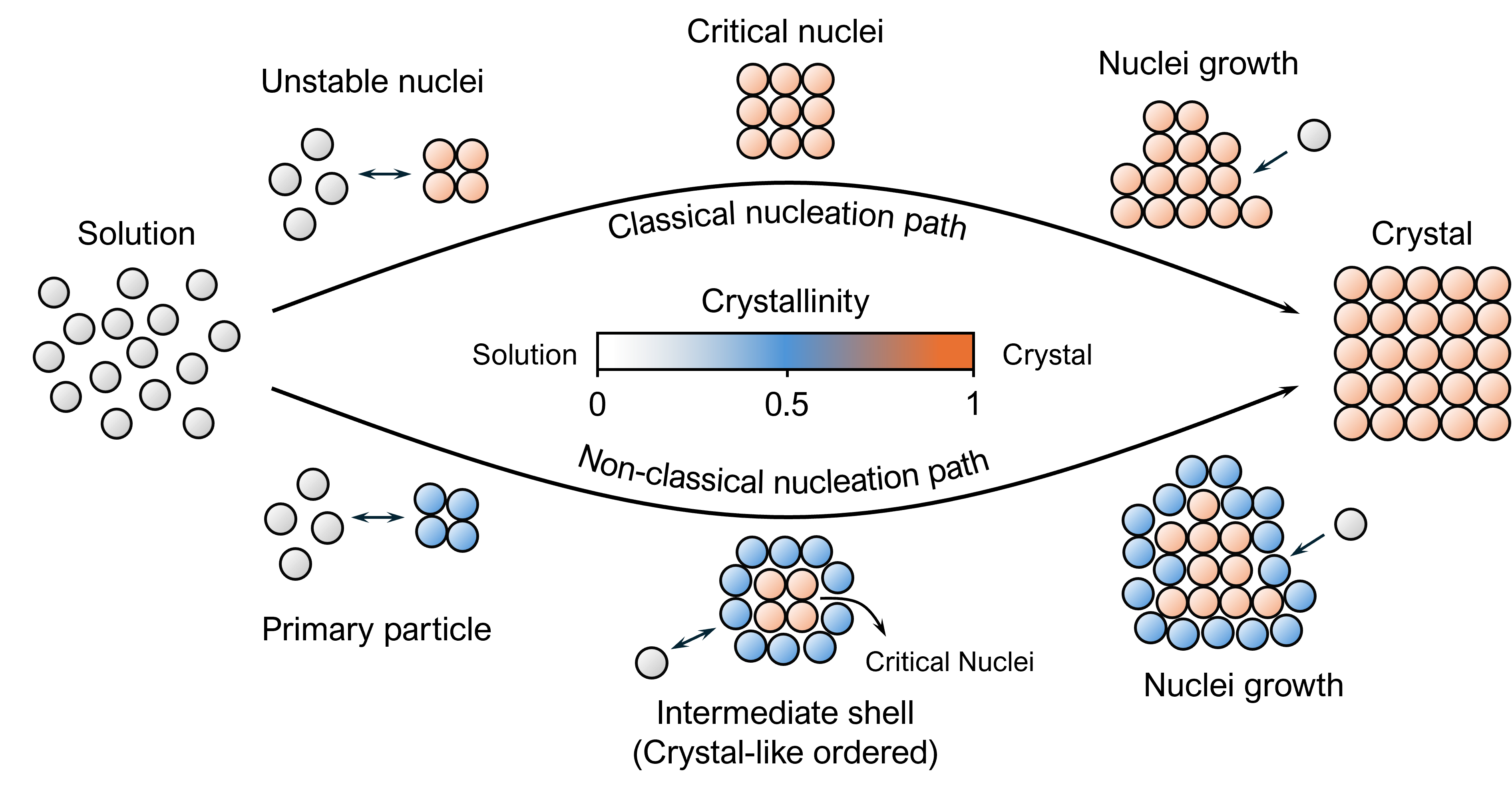}
  \caption{\small\textbf{Conceptual schematic of classical and non-classical nucleation pathways.} The classical pathway proceeds from unstable nuclei to a critical nucleus and crystal growth with a relatively sharp nucleus--parent interface. The non-classical pathway involves intermediate-shell formation around the nucleus during crystallization. Color denotes crystallinity.}
  \label{fig:conceptual-nucleation-pathways}
\end{figure}

\begin{figure}
  \centering
  \includegraphics[width=0.8\textwidth]{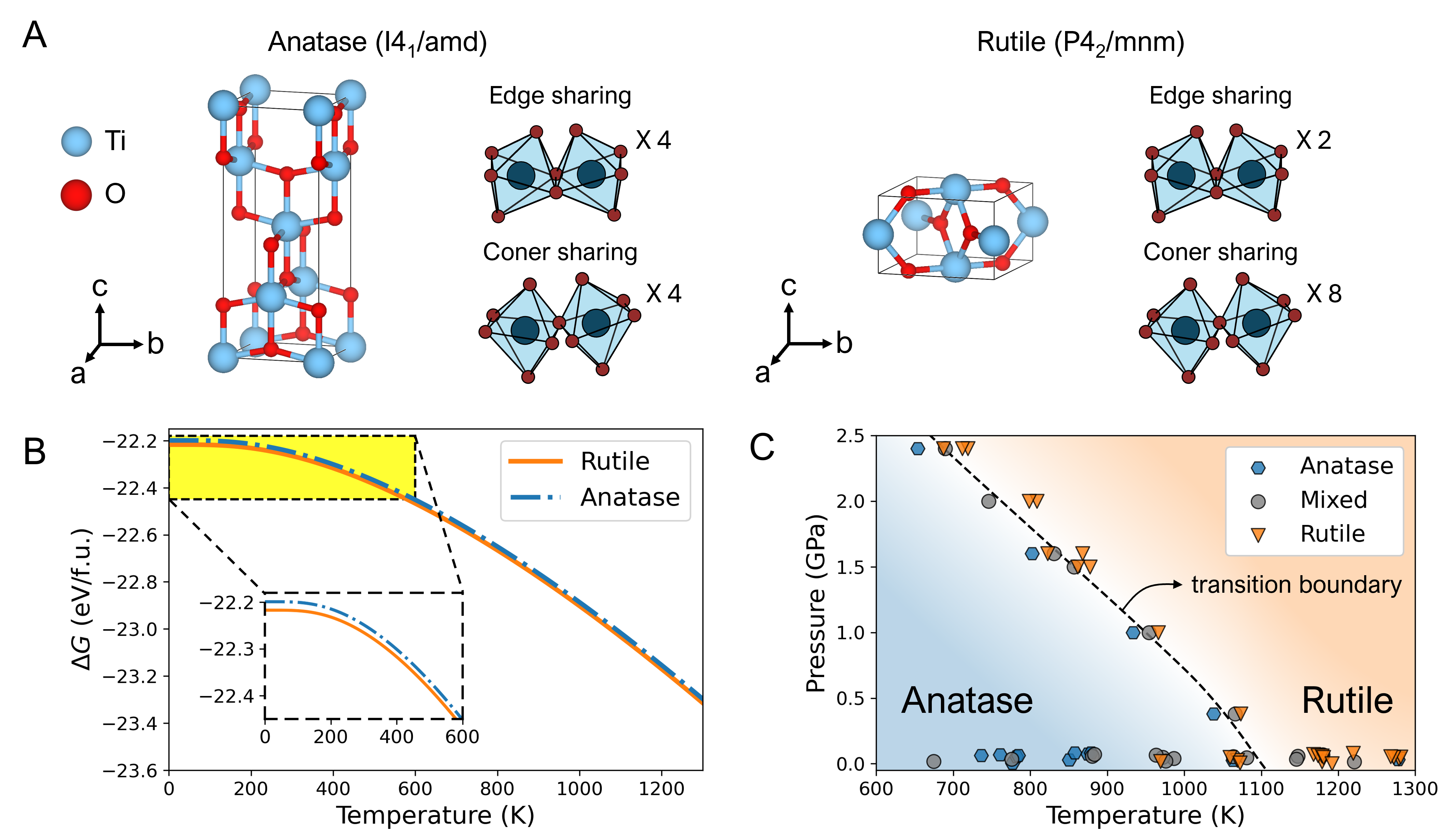}
  \caption{\small\textbf{Structural, thermodynamic, and experimental baseline for anatase--rutile selection.} (\textbf{A}) Crystal structures of anatase and rutile with TiO$_6$ motifs with different octahedral connectivity. (\textbf{B}) Calculated Gibbs free energies of anatase and rutile as a function of temperature; the inset shows an enlarged view of the highlighted region. (\textbf{C}) Reported experimental polymorph observations as a function of temperature and pressure. Blue hexagons, gray circles, and orange triangles denote anatase, mixed phase, and rutile, respectively; the dashed line marks the approximate transition boundary. Source data for the experimental observations are provided in Data S1.}
  \label{fig:anatase-rutile-baseline}
\end{figure}

\begin{figure}
  \centering
  \includegraphics[width=1.0\textwidth]{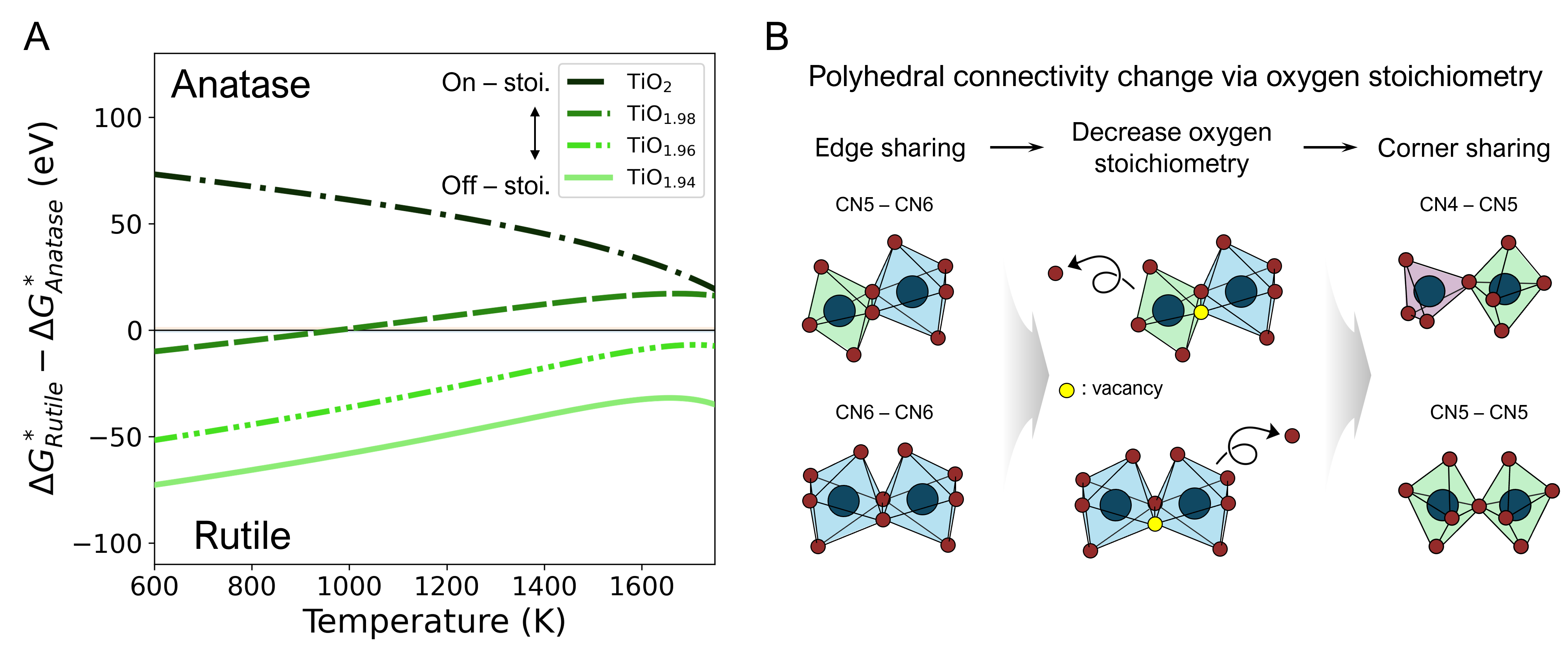}
  \caption{\small\textbf{Stoichiometry-dependent nucleation-barrier competition and local structural motif changes.} (\textbf{A}) Nucleation-barrier difference between rutile and anatase as a function of temperature across oxygen stoichiometries. Positive and negative values indicate anatase- and rutile-favored barrier ordering, respectively. (\textbf{B}) Schematic illustration of Ti--O polyhedral connectivity changes associated with decreasing oxygen stoichiometry, highlighting the shift from edge-sharing to corner-sharing motifs.}
  \label{fig:stoichiometry-barrier-motifs}
\end{figure}

\begin{figure}
  \centering
  \includegraphics[width=0.8\textwidth]{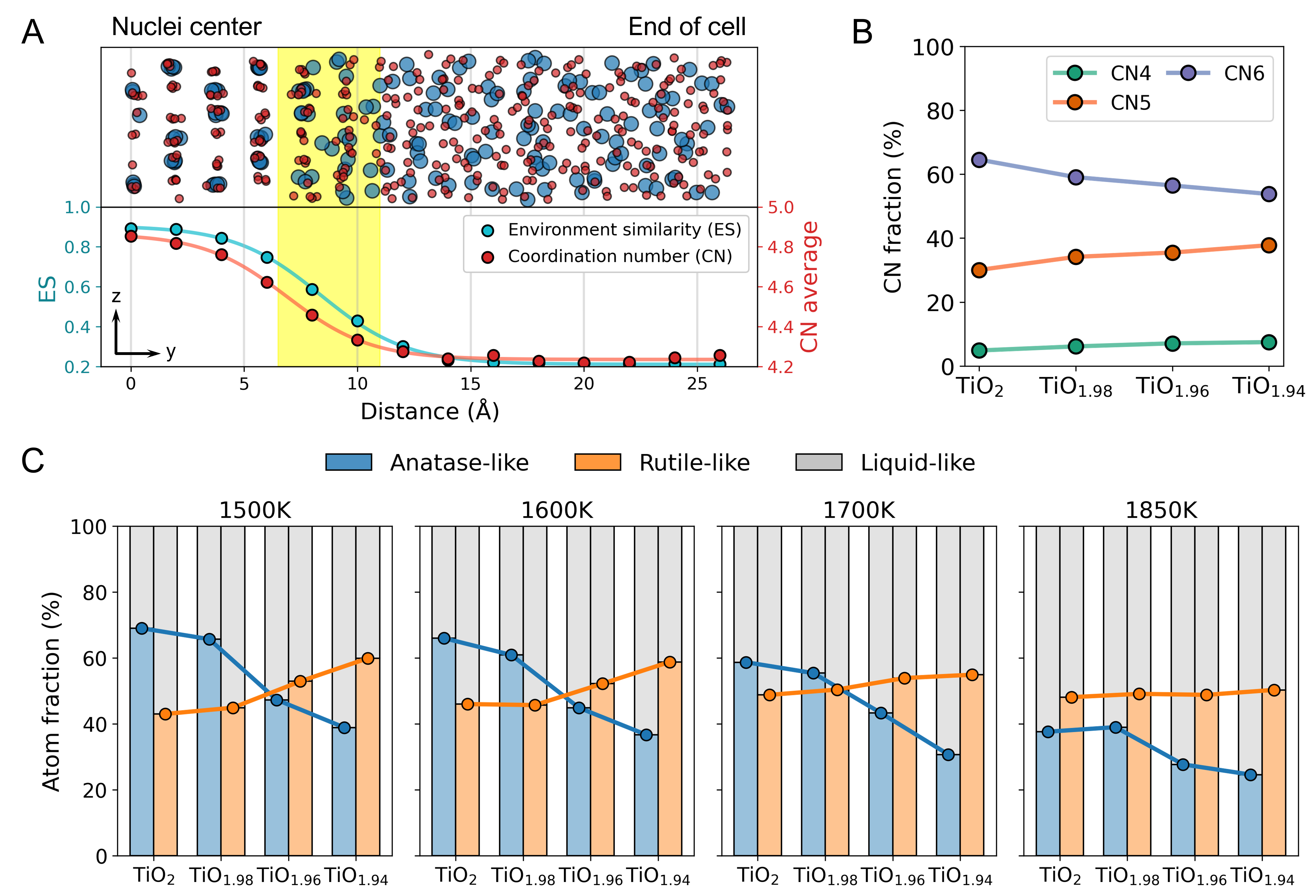}
  \caption{\small\textbf{Structural characteristics of the intermediate shell.} (\textbf{A}) Radial profiles of environment similarity (ES) and average Ti coordination number (CN) from the nucleus center toward the simulation cell edge for a representative anatase-seeded simulation at 1500~K. The yellow shaded region marks the intermediate shell. (\textbf{B}) Coordination-number distribution within the intermediate shell as a function of oxygen stoichiometry. (\textbf{C}) Polymorph-like and liquid-like atom fractions within the intermediate shell across oxygen stoichiometries and temperatures.}
  \label{fig:intermediate-shell-structure}
\end{figure}

\begin{figure}
  \centering
  \includegraphics[width=0.7\textwidth]{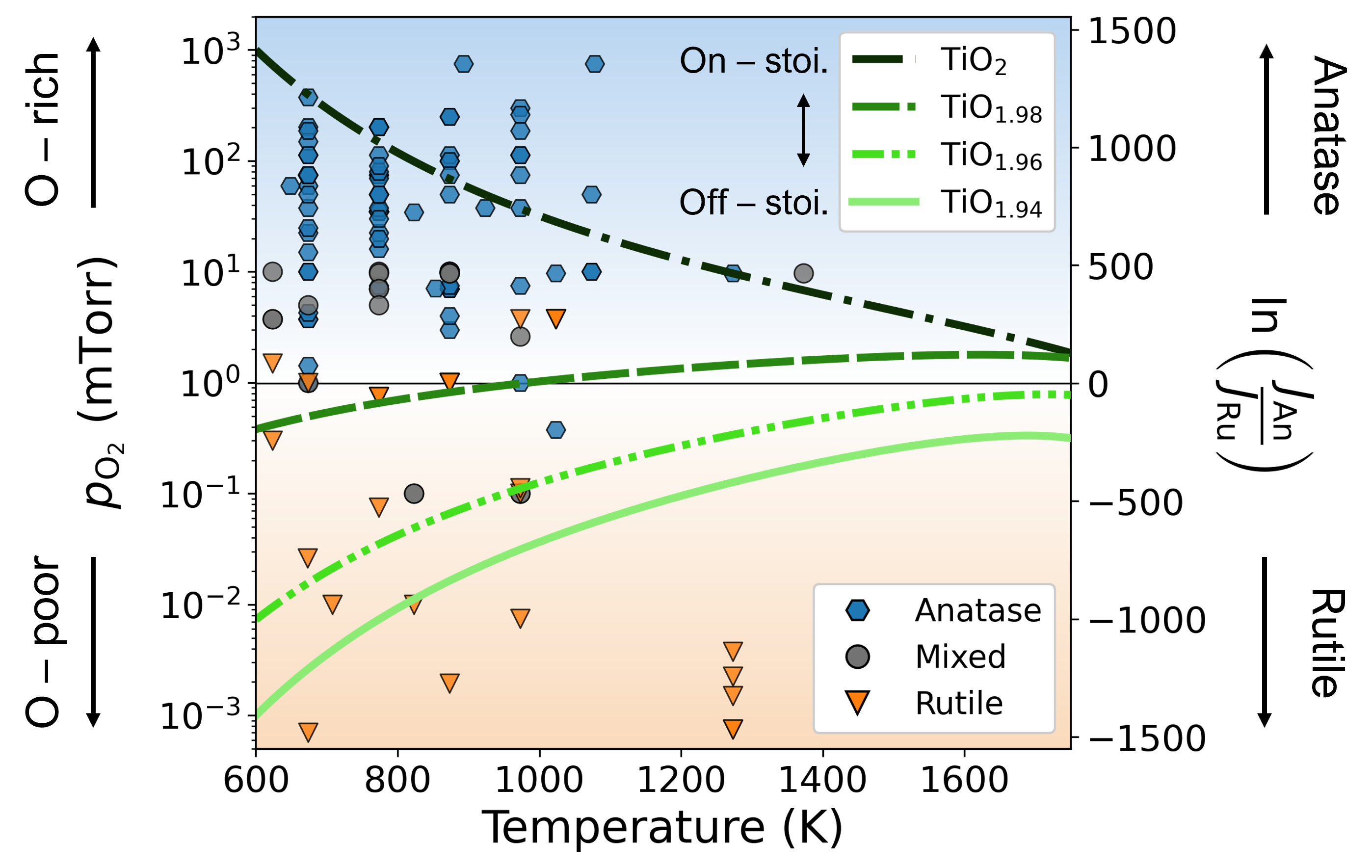}
  \caption{\small\textbf{Relative nucleation preference for anatase--rutile competition in TiO$_{2-x}$ compared with reported experimental phase observations.} The right y-axis shows the calculated relative nucleation preference across oxygen stoichiometries; positive and negative values indicate anatase- and rutile-favored nucleation, respectively. Experimental observations are shown on the left y-axis as a function of oxygen partial pressure ($p_{\mathrm{O_2}}$). Blue hexagons, gray circles, and orange triangles denote anatase, mixed phase, and rutile, respectively. Source data for the experimental observations are provided in Data S2.}
  \label{fig:relative-nucleation-preference}
\end{figure}

\begin{figure}
  \centering
  \includegraphics[width=1.0\textwidth]{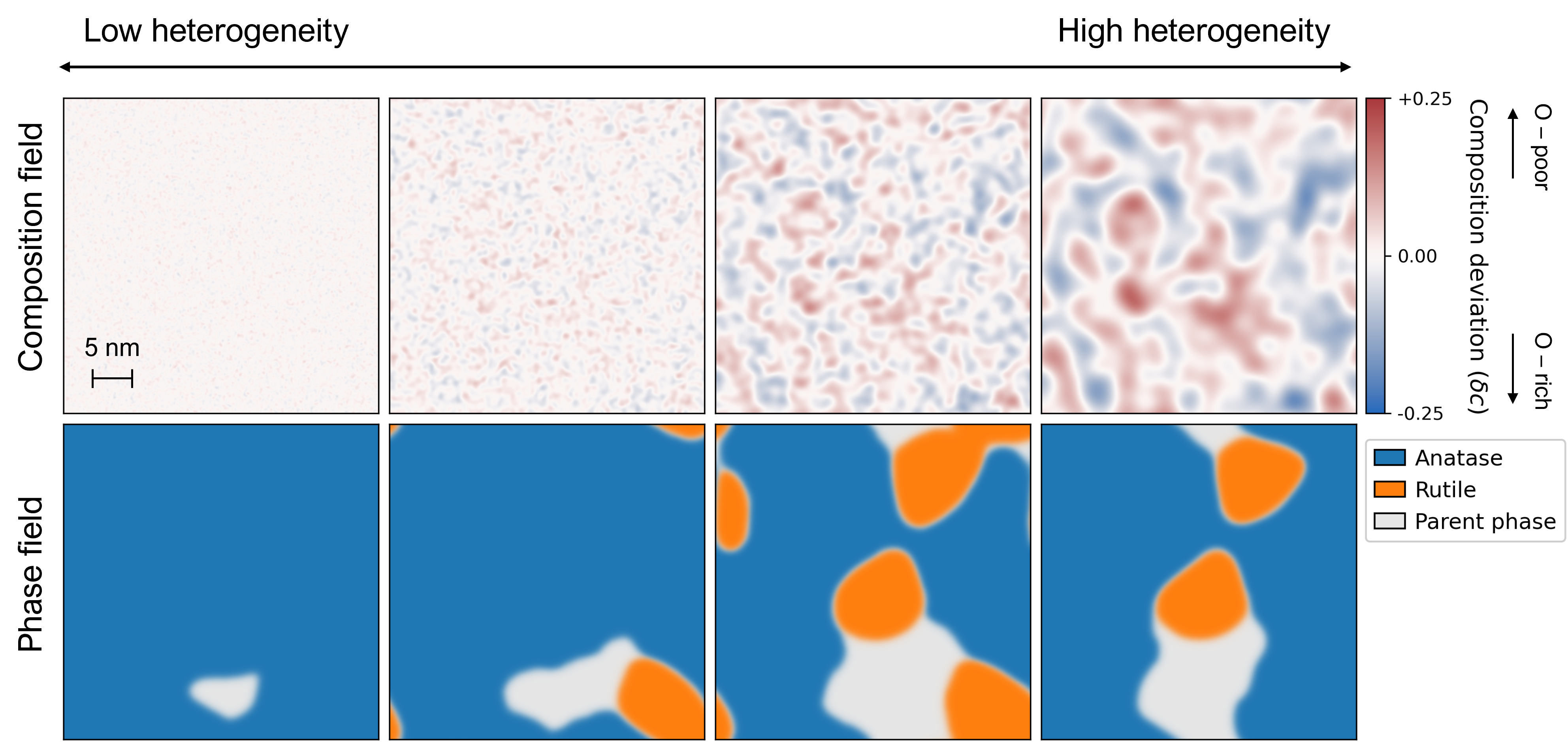}
  \caption{\small\textbf{Phase-field simulation from heterogeneous composition fields for the TiO$_2$ case.} The top row shows initial composition fields with increasing heterogeneity from left to right. The color scale shows the local deviation in normalized oxygen composition from its spatial mean ($\delta c$); positive (red) and negative (blue) values correspond to O-poor and O-rich local environments, respectively (see Supplementary Note 4 for details). The bottom row shows the corresponding phase fields after evolution, where blue, orange, and light gray denote anatase, rutile, and the parent phase, respectively. The 5~nm scale bar applies to all panels.}
  \label{fig:phase-field-heterogeneity}
\end{figure}

\clearpage
\bibliography{references}
\bibliographystyle{sciencemag}


\section*{Acknowledgments}
\paragraph*{Funding:}
\noindent This work was supported by the National Research Foundation of Korea (NRF) grant funded by the Ministry of Science and ICT (RS-2024-00444182 and RS-2026-25593368). The work at Washington University was partially supported by the National Science Foundation through award DMR-2145797.

\paragraph*{Author contributions:}
\noindent Conceptualization: H.U.L., R.M., and S.B.C. Methodology: H.U.L. Investigation: H.U.L., H.W.K., J.M.K. Visualization: H.U.L. Supervision: R.M. and S.B.C. Writing--original draft: H.U.L. Writing--review and editing: H.U.L., H.W.K., J.M.K., D.W.J., R.M. and S.B.C.

\paragraph*{Competing interests:}
\noindent The authors declare that they have no competing interests.

\paragraph*{Data, code and materials availability:}
\noindent The experimental source data are provided in Data S1 and S2, and all computational methods and models required to reproduce the results are fully described in this paper and the Supplementary Materials. In this study, no new materials were generated.


\subsection*{Supplementary materials}
Materials and Methods\\
Supplementary Text\\
Figures S1 to S9\\
Table S1\\
References \textit{(48--64)}\\
Data S1 to S2


\newpage
\def\CombinedSubmission{}
\ifdefined\CombinedSubmission
\let\FinishSupplement\relax
\else


\documentclass[12pt]{article}

\usepackage{newtxtext,newtxmath}
\usepackage{graphicx}
\usepackage[letterpaper,margin=1in]{geometry}
\linespread{1.5}
\frenchspacing

\renewenvironment{abstract}{\quotation}{\endquotation}
\date{}
\renewcommand\refname{References and Notes}

\makeatletter
\renewcommand{\fnum@figure}{\textbf{Figure \thefigure}}
\renewcommand{\fnum@table}{\textbf{Table \thetable}}
\makeatother


\usepackage{scicite}
\usepackage{url}


\def\scititle{Oxygen stoichiometry directs rutile–anatase phase selection through kinetic control of nucleation}
\newcommand{\sciauthorline}{Han Uk~Lee, Hyeon Woo~Kim, Ji Min~Kim, Dong Won~Jeon, Rohan~Mishra$^{\ast}$, Sung Beom~Cho$^{\ast}$}



\def\FinishSupplement{\end{document}}
\begin{document}

\fi

\renewcommand{\thefigure}{S\arabic{figure}}
\renewcommand{\thetable}{S\arabic{table}}
\renewcommand{\theequation}{S\arabic{equation}}
\renewcommand{\thepage}{S\arabic{page}}
\setcounter{figure}{0}
\setcounter{table}{0}
\setcounter{equation}{0}
\setcounter{page}{1}


\begin{center}
\section*{Supplementary Materials for\\ \scititle}

\sciauthorline\\
\small$^{\ast}$Corresponding author. Email: rmishra@wustl.edu; sungcho@skku.edu
\end{center}

\subsubsection*{This PDF file includes:}
Materials and Methods\\
Supplementary Text\\
Figures S1 to S9\\
Table S1\\
Captions for Data S1 to S2

\subsubsection*{Other Supplementary Materials for this manuscript:}
Data S1 to S2

\newpage


\subsection*{Materials and Methods}

\paragraph*{DFT calculations}
\noindent Density functional theory calculations were performed using the Vienna Ab initio Simulation Package (VASP) with the projector augmented-wave method and the generalized gradient approximation (GGA) in the Perdew--Burke--Ernzerhof (PBE) form~\cite{kresse1996efficient,perdew1996generalized,kresse1999ultrasoft}. Ti and O were described with the PAW\_PBE Ti\_sv and O potentials, respectively. The plane-wave basis set was expanded to a cutoff energy of 520~eV, and Gaussian smearing of 0.05~eV was used. Structural optimizations were truncated when the Hellmann--Feynman forces fell below 0.01~eV~\AA$^{-1}$, and the electronic energy was tightly converged to $1\times10^{-5}$~eV. The Brillouin zone was sampled with an automatically generated Gamma-centered $k$-point mesh at a reciprocal density of $100$/\AA$^{3}$ using pymatgen~\cite{ong2013python}. The GGA+$U$ approach was employed with an effective Hubbard parameter of $U=7$~eV applied to the Ti $3d$ states~\cite{curnan2015investigating}.

The anatase and rutile quasi-harmonic free energies were obtained from $2\times2\times2$ supercells of the conventional unit cells, sampled at 11 lattice-parameter scalings from $-5\%$ to $+5\%$ relative to the relaxed structure; at each fixed volume the internal coordinates were relaxed while the cell shape and volume were held fixed. Temperature-dependent free energies combined the volume-dependent static energies with vibrational free energies evaluated using PHONOPY~\cite{phonopy-phono3py-JPSJ} from VASP density-functional-perturbation-theory force constants computed in the same supercells, using a tighter convergence of $1\times10^{-8}$~eV and a $k$-point mesh of $3\times3\times1$ for anatase and up to $3\times3\times4$ for rutile.

For the grand-potential diagram (Fig.~\ref{fig:supp-grand-potential}), the same DFT settings were applied to stoichiometric Ti$_{32}$O$_{64}$ and single-vacancy Ti$_{32}$O$_{63}$ anatase and rutile cells (spin-polarized, ionic positions relaxed at fixed cell), to an isolated O$_2$ molecule used as the oxygen reference, and to the competing reduced phases (Ti$_2$O$_3$ and the Magnéli phases Ti$_3$O$_5$, Ti$_4$O$_7$, Ti$_5$O$_9$, and Ti$_6$O$_{11}$), which were fully relaxed.

\paragraph*{Molecular dynamics}
\noindent Molecular dynamics simulations were performed with LAMMPS~\cite{thompson2022lammps} using the Matsui--Akaogi (MA) Ti--O interatomic potential~\cite{matsui1991molecular}. This potential has been used to model the crystallization of amorphous TiO$_2$ and the anatase-to-rutile transformation in molecular dynamics~\cite{zhang2023investigating}. The short-range interactions were described by Buckingham terms with a 12~\AA{} cutoff, and long-range Coulomb interactions were treated with PPPM using a relative accuracy of $10^{-6}$. Ti and O atoms were assigned partial charges of $+2.196e$ and $-1.098e$, respectively. Stoichiometric metadynamics cells contained 144 Ti and 288 O atoms, giving 432 atoms in total; for oxygen-deficient compositions, the Ti count was fixed and O atoms were removed to set the target stoichiometry.

All MD simulations used a 1~fs timestep. After energy minimization, velocities were initialized at the target temperature. The systems were first propagated for 10~ps in the NVE ensemble to remove initial transients and then equilibrated for 30~ps in the NPT ensemble at 1~atm before production. The NPT ensemble was implemented using an NPH barostat together with a CSVR thermostat~\cite{bussi2007canonical}; the thermostat and barostat damping times were 1~ps during production. Production metadynamics simulations were performed for 50~ns with PLUMED coupled to LAMMPS in the same NPT ensemble.

\paragraph*{Metadynamics and collective variable}
\noindent Metadynamics simulations were performed at 2300, 2375, 2450, 2575, and 2700~K for TiO$_2$, TiO$_{1.98}$, TiO$_{1.96}$, and TiO$_{1.94}$ with PLUMED~\cite{plumed2019promoting,tribello2014plumed} using the Ti-centered Environment Similarity (ES)~\cite{piaggi2019calculation} described in Supplementary Note 2. The kernel for anatase was defined using $\sigma=0.05$, with the reference Ti environment corresponding to the 8 nearest Ti neighbors of the anatase lattice. The kernel for rutile was defined using $\sigma=0.04$, with the reference Ti environment corresponding to the 22 nearest Ti neighbors of the rutile lattice. These neighbor counts define the Ti-neighbor shell included in each polymorph-specific reference environment. The per-atom similarities were converted into a smooth Ti-centered ES count using $k_0=0.5$, $p=6$, and $q=12$; this count was used to bias the growth of target-polymorph-like Ti environments.

Well-tempered metadynamics (WTMetaD) with adaptive Gaussian widths~\cite{branduardi2012metadynamics} was used for the nucleation simulations. The bias factors for anatase and rutile were 120 and 300, respectively, and the initial Gaussian height parameter was 100. The adaptive Gaussian width was bounded between 2 and 20 for both anatase and rutile. The metadynamics bias was deposited every 500 MD steps, corresponding to 0.5~ps, giving up to $1.0\times10^5$ Gaussian depositions over each 50~ns trajectory. Reweighted free-energy surfaces were reconstructed from the WTMetaD trajectories. For the driving-force analysis, the parent-like ($\Omega_\ell$) and polymorph-like ($\Omega_i$) basins were divided at the midpoint of the ES count, $s_{\mathrm{ES}}=72$, i.e.\ half of the 144-Ti saturation, so that configurations with more than half of the Ti environments polymorph-like ($s_{\mathrm{ES}}\ge72$) were assigned to the polymorph basin.

As the crystal can form in orientations misaligned with the simulation box during the MD simulation, an additional quantity was introduced to suppress this~\cite{piaggi2019calculation}. A global Steinhardt $Q_6$ order parameter was evaluated over all atoms using a cubic switching function with $D_0=0.21$ and $D_\mathrm{max}=0.25$, and compared with the Ti-centered ES count through
\begin{equation}
\Delta_{Q_6-\mathrm{ES}} =
\frac{Q_6-Q_6^{l}}{Q_6^{s}-Q_6^{l}}
-
\frac{s_{\mathrm{ES}}-s_{\mathrm{ES}}^{l}}{s_{\mathrm{ES}}^{s}-s_{\mathrm{ES}}^{l}},
\end{equation}
where $s_{\mathrm{ES}}$ is the Ti-centered ES count. Because both terms are normalized between their parent-phase-like and fully crystalline reference values, $\Delta_{Q_6-\mathrm{ES}}$ stays close to zero only when the global order $Q_6$ and the ES count $s_{\mathrm{ES}}$ grow together; a value far from zero therefore signals crystallization in an orientation differing from the target. The parameters were $Q_6^{l}=0.02$, $s_{\mathrm{ES}}^{l}=0$, and $s_{\mathrm{ES}}^{s}=144$ for both polymorphs, where $s_{\mathrm{ES}}^{s}$ corresponds to the fully crystalline saturation over the 144 Ti atoms in the metadynamics cell. The solid-reference values were $Q_6^{s}=0.22$ for anatase and $Q_6^{s}=0.27$ for rutile. To keep this mismatch coordinate below a prescribed threshold, the following upper-wall potential was applied:
\begin{equation}
V(\Delta_{Q_6-\mathrm{ES}})=
\begin{cases}
\kappa_i \left(\Delta_{Q_6-\mathrm{ES}}-\Delta_i^0\right)^{m_i}, & \Delta_{Q_6-\mathrm{ES}} > \Delta_i^0,\\
0, & \Delta_{Q_6-\mathrm{ES}} \le \Delta_i^0,
\end{cases}
\end{equation}
where $\Delta_i^0$ is the wall position and $m_i$ is the wall exponent. The parameters were $\Delta_i^0=0.25$, $\kappa_i=10^5$, and $m_i=2$ for anatase, and $\Delta_i^0=0.12$, $\kappa_i=3\times10^5$, and $m_i=4$ for rutile. This wall restrained off-target orientational growth; the metadynamics bias itself was applied only to the Ti-centered ES coordinate.

\paragraph*{Seeding simulations}
\noindent Motivated by previous approaches~\cite{espinosa2016seeding}, seeding simulations for anatase and rutile were performed at 1500, 1600, 1700, 1850, and 2000~K in LAMMPS using the same MA potential, timestep, pressure control, and thermostatting scheme as described above. The initial anatase and rutile configurations were replicated by $14\times14\times6$ and $9\times9\times14$, respectively, giving stoichiometric simulation cells containing 14,112 atoms for anatase and 6,804 atoms for rutile before oxygen removal. For each temperature and oxygen stoichiometry, a spherical crystalline seed was defined at the center of the box; the seed radius was varied across conditions rather than fixed to a single value. The parent matrix surrounding the spherical seed was thermally disordered. When oxygen-deficient compositions were considered, O atoms were removed from the parent-region atoms to set the target stoichiometry.

The parent-region atoms were heated to 5000~K for 50~ps and then cooled back to the target temperature for 10~ps, while the seed region was retained. Before the production seeding trajectory, the full system was equilibrated at the target temperature and 1~atm. During a 10~ps interface-equilibration stage, atoms in the seed region were restrained with a harmonic self-spring to preserve the initial seed structure while the surrounding matrix relaxed. The restraints were then removed, and the production seeding trajectory was propagated for 500~ps in the NPT ensemble.

\paragraph*{Phase-field simulations}
\noindent Phase-field simulations for Fig.~6 and Figs.~S7--S9 were performed using in-house code. The simulations used two nonconserved order parameters for anatase-like and rutile-like domains and a conserved normalized composition field $c$. The phase-field free-energy functional was evaluated with composition-dependent thermodynamic inputs at 1000~K. The order parameters and composition field were evolved by Allen--Cahn and Cahn--Hilliard dynamics, respectively, using a semi-implicit Fourier-spectral scheme~\cite{allen1979microscopic,cahn1958free,chen1998applications}. Details of the phase-field formulation and numerical implementation are provided in Supplementary Note 4.

The phase-field calculations were carried out on a two-dimensional $320~\text{\AA}\times320~\text{\AA}$ domain using a $516\times516$ grid. The initial composition field was generated as a Fourier-correlated random texture. The spatial mean $c_0$ was set to 0, 0.02, 0.04, and 0.06 for TiO$_2$, TiO$_{1.98}$, TiO$_{1.96}$, and TiO$_{1.94}$, respectively. Positive and negative values of the local deviation $\delta c(\mathbf{x})$, defined in Supplementary Note 4.1, correspond to locally O-poor and O-rich environments, respectively. For each stoichiometry, four composition fields were used. Each random texture was generated by applying a Gaussian filter in Fourier space to Gaussian white noise, normalizing the filtered field to unit spatial standard deviation, and rescaling it to the prescribed fluctuation amplitude. The field was then centered so that $\langle\delta c\rangle=0$, thereby preserving $c_0$. For all stoichiometries, the fluctuation amplitudes were 0.010, 0.020, 0.040, and 0.060, paired in order with characteristic texture lengths of approximately 6, 16, 31, and 62~\AA. Both parameters were increased together from left to right, producing progressively stronger and coarser composition variations.

Each calculation was initialized with 10 seed sites whose locations were selected independently of the CFM, subject to boundary-exclusion and non-overlap constraints. Each seed identity was sampled from the CFM-derived probabilities evaluated at its local initial composition. These two-dimensional calculations qualitatively illustrate polymorph selection during growth.


\subsection*{Supplementary Text}

\subsubsection*{Supplementary Note 1 -- Equilibrium stability of anatase and rutile in TiO$_{2-x}$}

\begin{figure}
  \centering
  \includegraphics[width=0.7\textwidth]{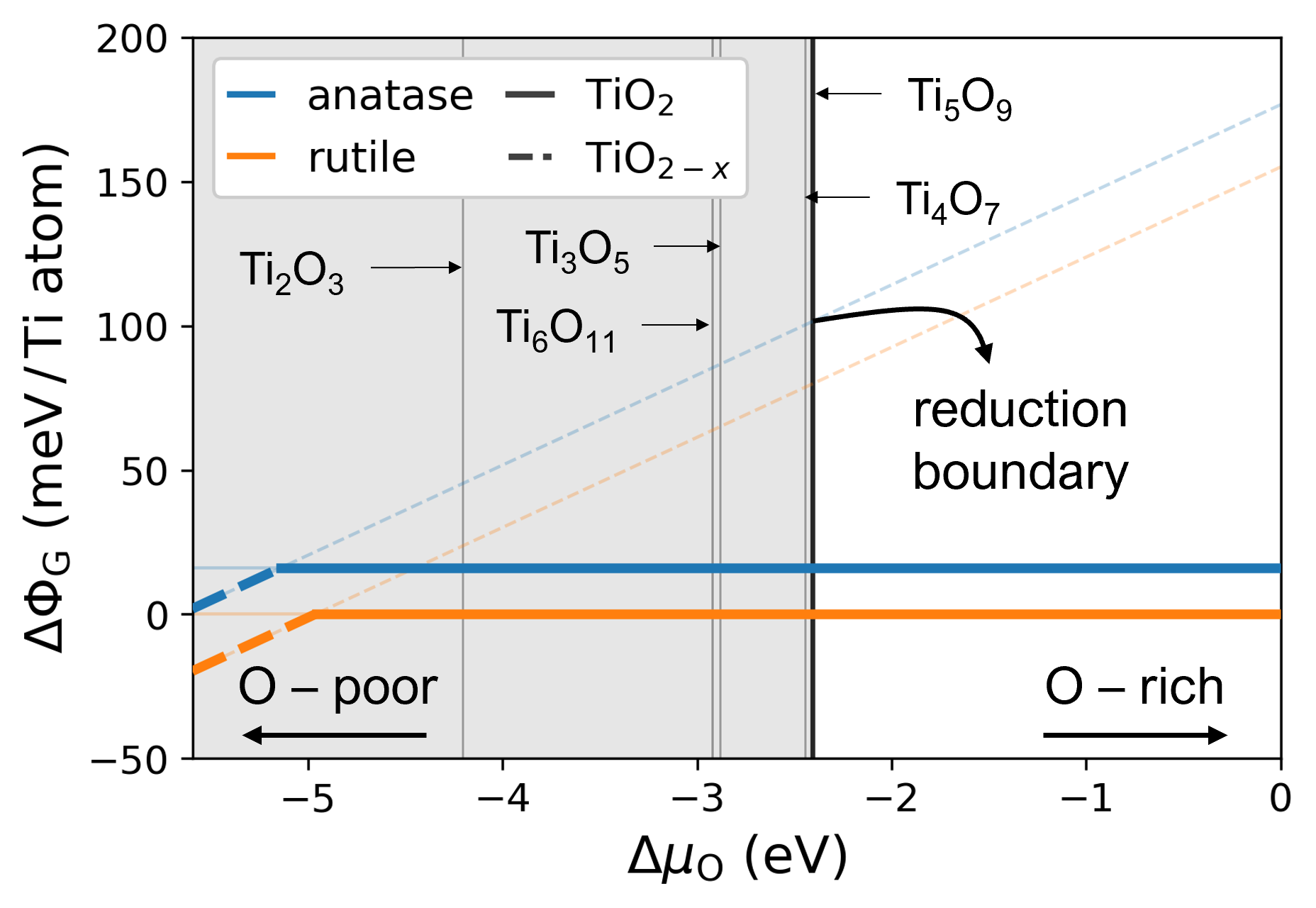}
  \caption{\small\textbf{Relative grand-potential diagram of the TiO$_{2-x}$ system.} Relative grand potential per Ti atom, $\Delta\Phi_\mathrm{G}$, for anatase (blue) and rutile (orange) versus oxygen chemical potential $\Delta\mu_\mathrm{O}$, referenced to stoichiometric rutile TiO$_2$ ($\Delta\Phi_\mathrm{G}=0$). Solid and dashed curves denote the stoichiometric (TiO$_2$) and oxygen-deficient (TiO$_{2-x}$) cells; the bold curve traces the lower envelope. The shaded region lies below the reduction boundary, where TiO$_2$ is unstable against reduced Ti--O phases (vertical lines).}
  \label{fig:supp-grand-potential}
\end{figure}

As discussed in the main text, oxygen stoichiometry affects TiO$_2$ polymorph selection. To test whether this dependence can be explained by equilibrium thermodynamics, we evaluate the grand potential
\begin{equation}
\Phi_\mathrm{G}=\frac{E_\mathrm{tot}-N_\mathrm{O}\mu_\mathrm{O}}{N_\mathrm{Ti}},
\label{eq:grand-potential}
\end{equation}
\noindent
where $E_\mathrm{tot}$ is the total energy, $N_\mathrm{O}$ and $N_\mathrm{Ti}$ are the numbers of oxygen and titanium atoms, and $\mu_\mathrm{O}$ is the oxygen chemical potential. We report values relative to rutile TiO$_2$, the thermodynamically stable polymorph,
\begin{equation}
\Delta\Phi_\mathrm{G}=\Phi_\mathrm{G}-\Phi_\mathrm{G}^{\mathrm{rutile}},
\label{eq:relative-grand-potential}
\end{equation}
\noindent
plotted in Fig.~\ref{fig:supp-grand-potential} against $\Delta\mu_\mathrm{O}$, the oxygen chemical potential relative to the oxygen-rich limit.

Because oxygen-poor conditions can also drive decomposition of TiO$_2$ into reduced Ti--O phases, we bound the chemical-potential range over which TiO$_2$ remains stable. We evaluate the decomposition equilibrium of rutile TiO$_2$ with each competing reduced phase (Ti$_2$O$_3$ and the Magnéli phases Ti$_3$O$_5$, Ti$_4$O$_7$, Ti$_5$O$_9$, and Ti$_6$O$_{11}$); the most competitive of these, Ti$_5$O$_9$, sets the highest oxygen chemical potential at which TiO$_2$ decomposes, and thus the O-poor limit of the stability window (shaded region in Fig.~\ref{fig:supp-grand-potential}). Even under oxygen-rich conditions, rutile is thermodynamically preferred over anatase, and increasing oxygen deficiency does not bring anatase below rutile; instead, it ultimately drives TiO$_2$ to decompose into the reduced phases.

\subsubsection*{Supplementary Note 2 -- Nucleation behavior of TiO$_{2-x}$}

This note describes the computational workflow used to evaluate nucleation behavior in TiO$_{2-x}$. In the main manuscript, the barrier component of this behavior is summarized by the barrier difference $\Delta G^*_\mathrm{rutile}-\Delta G^*_\mathrm{anatase}$, which indicates whether anatase or rutile nucleation is more accessible at a given temperature and oxygen stoichiometry. Here, we provide the technical details behind this calculation: the derivation of the critical nucleation barrier, the enhanced-sampling procedure used to obtain the thermodynamic driving force, and the seeding procedure used to determine the critical nucleus size.

The note is organized as follows. Section 2.1 derives the critical nucleation barrier from a cluster free-energy expression. Section 2.2 explains how metadynamics is used to sample rare crystallization events, how Environment Similarity is used to define the sampling coordinate, and how the reconstructed free-energy surfaces are converted into parent-to-polymorph driving forces. Section 2.3 explains how the critical nucleus size is determined from seeding simulations. Finally, Section 2.4 shows how these quantities are combined to construct the barrier difference shown in Fig. 3A of the main manuscript.

\paragraph*{2.1 Critical nucleation barrier}

Nucleation of a crystalline polymorph can be viewed as the formation of a small solid-like cluster within a metastable parent phase. The free energy of this cluster is governed by two competing contributions. The bulk term favors growth because transforming the parent phase into the solid polymorph lowers the free energy by the chemical-potential driving force. In contrast, the interfacial term penalizes cluster formation because a new boundary must be created between the nucleus and the surrounding parent phase.

As a result, very small clusters are unstable and tend to dissolve, because the interfacial penalty dominates. Once the cluster becomes sufficiently large, the bulk driving force dominates and the cluster can continue to grow. The maximum of this free-energy profile defines the critical nucleus, and the corresponding free-energy maximum is the critical nucleation barrier. Comparing the barriers for anatase and rutile indicates which nucleation pathway has the lower barrier at a given temperature and oxygen stoichiometry.

The critical nucleation barrier is approximated as
\begin{equation}
\Delta G_i^*
\simeq
\frac{1}{2}
\left|\Delta\mu_{\ell\to i}\right|n_{c,i},
\label{eq:critical-barrier-compact-intro}
\end{equation}
\noindent\ 
where $|\Delta\mu_{\ell\to i}|$ is the chemical-potential difference between the parent phase and polymorph $i$, and $n_{c,i}$ is the critical nucleus size for polymorph $i$. Therefore, the nucleation barrier can be evaluated once these two quantities are determined. For polymorph $i$, the nucleation free energy as a function of nucleus radius is
\begin{equation}
\Delta G_i(r)
=
-\frac{4\pi r^3}{3v_i}\left|\Delta\mu_{\ell\to i}\right|
+4\pi r^2\gamma_i,
\label{eq:cluster-free-energy-radius}
\end{equation}
\noindent\ 
where $r$ is the nucleus radius, $v_i$ is the molecular volume of polymorph $i$, $|\Delta\mu_{\ell\to i}|$ is the chemical-potential difference between the parent phase and the solid polymorph, and $\gamma_i$ is the effective nucleus--parent interfacial energy.

Convert the radius-based expression to a cluster-size expression using
\begin{equation}
n=\frac{4\pi r^3}{3v_i}.
\label{eq:cluster-size-radius}
\end{equation}
\noindent\ 
Then the free energy can be written as
\begin{equation}
\Delta G_i(n)
=
-n\left|\Delta\mu_{\ell\to i}\right|
+c_i\gamma_i n^{2/3},
\label{eq:cluster-free-energy-size}
\end{equation}
\noindent\ 
where $c_i$ is a shape and molecular-volume factor. Maximizing $\Delta G_i(n)$ gives the critical size:
\begin{equation}
n_{c,i}
=
\left(
\frac{2c_i\gamma_i}{3\left|\Delta\mu_{\ell\to i}\right|}
\right)^3.
\label{eq:critical-size-cluster}
\end{equation}
\noindent\ 
This gives the scaling:
\begin{equation}
\Delta G_i^*
\propto
\frac{\gamma_i^3}{\left|\Delta\mu_{\ell\to i}\right|^2}.
\label{eq:supp-barrier-scaling}
\end{equation}
\noindent\ 
At the critical size, the surface term satisfies
\begin{equation}
c_i\gamma_i n_{c,i}^{2/3}
=
\frac{3}{2}
\left|\Delta\mu_{\ell\to i}\right|n_{c,i}.
\label{eq:critical-surface-balance}
\end{equation}
\noindent\ 
Substitution gives the form used in the manuscript:
\begin{equation}
\Delta G_i^*
\simeq
\frac{1}{2}
\left|\Delta\mu_{\ell\to i}\right|n_{c,i},
\qquad
i\in\{\mathrm{anatase},\mathrm{rutile}\}.
\label{eq:critical-barrier-compact}
\end{equation}
\noindent\ 
The following two sections explain how each quantity is obtained.

\paragraph*{2.2 Enhanced sampling for rare events}

$|\Delta\mu_{\ell\to i}|$ is the chemical-potential difference between the parent phase and polymorph $i$. It represents the free-energy driving force for the transformation from the parent phase to polymorph $i$ at a given temperature and oxygen stoichiometry. To evaluate this quantity in simulations, the transformation pathway must be sampled so that the free-energy separation between parent-phase-like and crystal-like states can be measured.

However, sampling such transformations has long been a central challenge in molecular dynamics studies of nucleation and crystallization. These processes occur as rare events, and direct unbiased trajectories can remain trapped in one structural state over accessible simulation timescales. Enhanced sampling is therefore required. In this work, we use metadynamics to promote rare parent-to-polymorph transformations by applying a bias along a crystallization coordinate.

\subparagraph*{2.2.1 Metadynamics}

Metadynamics enhances sampling by adding a history-dependent bias along a selected coordinate. Schematically, the accumulated bias can be written as a sum of Gaussian functions deposited along the trajectory:
\begin{equation}
V(s,t)
=
\sum_{t'<t}
W(t')
\exp\left[
-\frac{\left(s-s(t')\right)^2}{2\sigma_s^2}
\right],
\label{eq:metadynamics-bias}
\end{equation}
\noindent\ 
where $s$ is the biased coordinate, $W(t')$ is the height of the Gaussian deposited at time $t'$, $\sigma_s$ is its width, and $t$ is the current simulation time; the sum runs over all earlier deposition times $t'<t$. Each Gaussian is deposited at $s(t')$, the coordinate value occupied by the system at time $t'$, so the bias accumulates in regions previously visited by the trajectory and progressively fills the corresponding free-energy wells. As a well fills, the system is no longer held there and is pushed into coordinate regions it has not yet visited; iterating this lets the trajectory escape long-lived states and explore a broader coordinate range instead of staying trapped in one basin. Once deposition has roughly leveled the landscape, the accumulated bias mirrors the underlying free-energy surface, which is recovered from it. We use well-tempered metadynamics (WTMetaD) with adaptive Gaussian widths: $\sigma_s$ is adjusted on the fly to the local fluctuations of the coordinate, and the deposited height is progressively reduced on revisiting a region, so the bias converges smoothly to a stationary free-energy estimate rather than overfilling the wells. For a detailed theoretical treatment of metadynamics, see Refs.~\cite{laio2002escaping,bussi2020using}.

Because the bias is applied along $s$, the choice of this coordinate is central. For crystallization, $s$ must track the structural progress from a parent-phase-like state toward crystal-like order. Here, $s$ is constructed from Environment Similarity (ES) to track the parent-to-polymorph transformation and evaluate the corresponding free-energy landscape.

\subparagraph*{2.2.2 Environment Similarity}

Environment Similarity (ES) quantifies how closely the local atomic arrangement around each Ti atom resembles a reference crystalline environment~\cite{piaggi2019calculation}. Each Ti-centered environment is represented by a cloud of neighboring atoms, which is compared with the corresponding cloud in a reference polymorph. The overlap is converted into a similarity score that approaches one when the local environment closely matches the reference and decreases for disordered or differently ordered environments. Because anatase, rutile, and the disordered parent phase have distinct local arrangements around Ti, ES can distinguish parent-phase-like from polymorph-like environments while tracking the structural progression between them.

\begin{figure}
  \centering
  \includegraphics[width=0.9\textwidth]{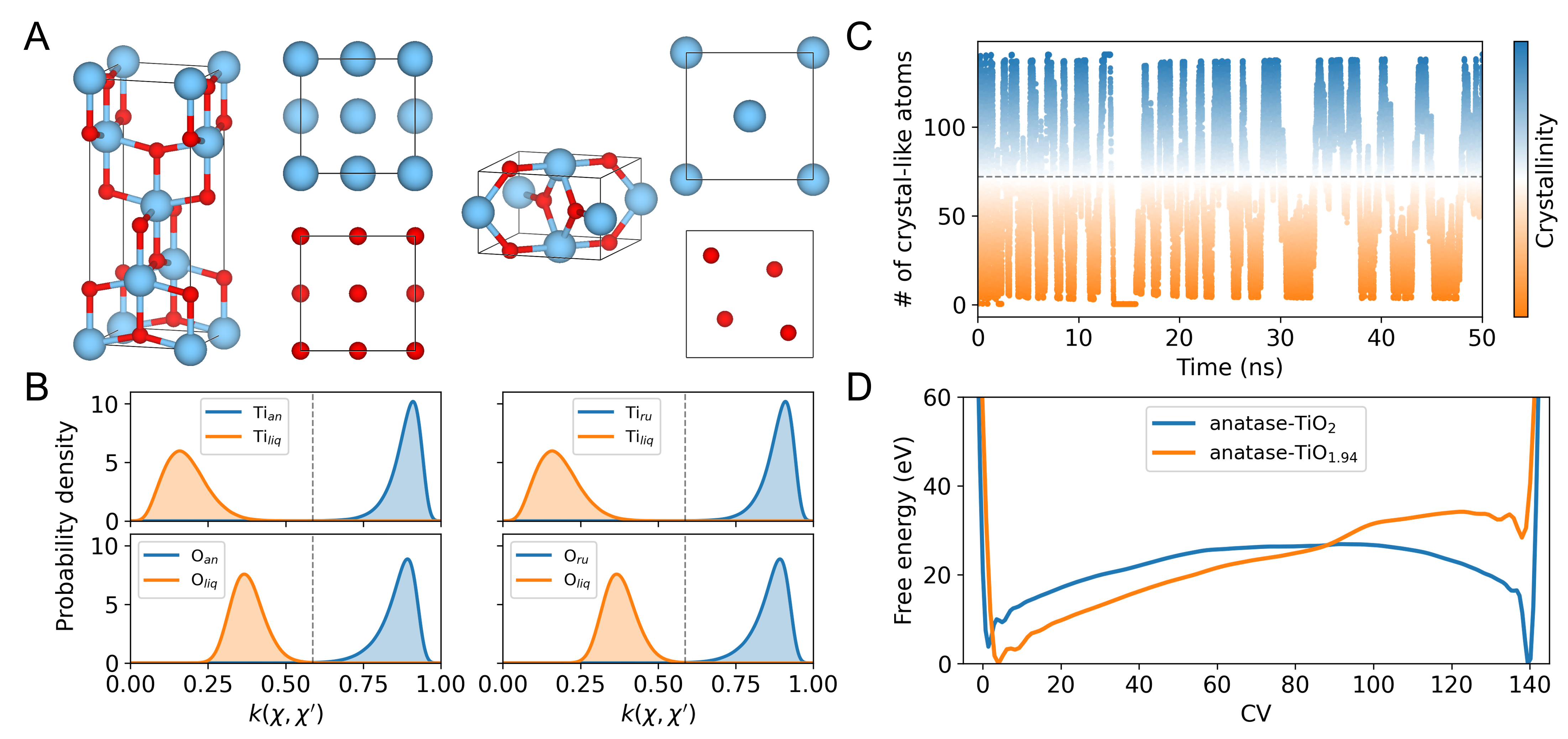}
  \caption{\small\textbf{Collective variable and metadynamics results.} \textbf{(A)} Anatase and rutile reference structures and the Ti and O sublattices used to define the ES kernel. \textbf{(B)} ES kernel distributions for crystalline and parent environments, showing separation of the relevant structural states. \textbf{(C)} Time evolution of the ES-based crystallinity coordinate during metadynamics, showing repeated sampling across the relevant coordinate range; the color bar denotes crystallinity. \textbf{(D)} Representative free-energy surfaces reconstructed along the ES coordinate for anatase TiO$_2$ and TiO$_{1.94}$ at 2450 K; blue and orange curves correspond to TiO$_2$ and TiO$_{1.94}$, respectively.}
  \label{fig:supp-metadynamics-es}
\end{figure}

Figure~\ref{fig:supp-metadynamics-es}A shows the Ti-centered reference environments $\chi_\mathrm{an}$ and $\chi_\mathrm{ru}$, constructed from bulk anatase (I4$_1$/amd) and rutile (P4$_2$/mnm), respectively.

A local Ti environment $\chi$ is compared with a reference environment $\chi_0$ through a normalized kernel that measures the overlap of their neighbor densities,
\begin{equation}
\tilde{k}_{\chi_0}(\chi)
=
\frac{k_{\chi_0}(\chi)}{k_{\chi_0}(\chi_0)},
\qquad
k_{\chi_0}(\chi)
=
\sum_{a\in\chi}
\sum_{b\in\chi_0}
\exp\left[
-\frac{\left|\mathbf{r}_a-\mathbf{r}^{0}_b\right|^2}{4\sigma^2}
\right],
\label{eq:normalized-es-kernel}
\end{equation}
\noindent\ 
where $\mathbf{r}_a$ and $\mathbf{r}^{0}_b$ are neighbor positions in the sampled and reference environments and $\sigma$ sets the spatial resolution. Each term compares one neighbor in the sampled environment with one in the reference, so the double sum is large when the two neighbor clouds overlap closely. The normalization gives $\tilde{k}_{\chi_0}(\chi)=1$ for an environment identical to the reference and a value approaching zero as the local structure departs from it. Comparing each Ti environment with the two references therefore assigns similarity scores $\tilde{k}_{\chi_\mathrm{an}}(\chi)$ and $\tilde{k}_{\chi_\mathrm{ru}}(\chi)$ that distinguish parent-phase-like from polymorph-like order. The resulting distributions separate the relevant states (Fig.~\ref{fig:supp-metadynamics-es}B), confirming that ES distinguishes the structural states required for crystallization sampling.

These per-atom scores are aggregated into a global crystallinity coordinate that serves as the collective variable (CV) biased in metadynamics. For a polymorph reference $\chi_0\in\{\chi_\mathrm{an},\chi_\mathrm{ru}\}$, this coordinate provides a smooth count of Ti environments resembling that polymorph:
\begin{equation}
s
=
\sum_{\alpha=1}^{N}
\left(
1
-
\frac{1-\left(\tilde{k}_{\chi_0}(\chi_\alpha)/k_0\right)^{p}}
{1-\left(\tilde{k}_{\chi_0}(\chi_\alpha)/k_0\right)^{q}}
\right),
\label{eq:crystallinity-coordinate}
\end{equation}
\noindent\ 
where the sum runs over all $N$ Ti atoms and $\chi_\alpha$ is the environment of Ti atom $\alpha$. The parameter $k_0$ is a threshold on the normalized similarity ($0<k_0<1$), such that $\tilde{k}_{\chi_0}(\chi_\alpha)/k_0$ exceeds one for polymorph-like atoms and falls below one for parent-phase-like atoms. The integer exponents $p$ and $q$ control the sharpness of the transition around $k_0$. Each summand is close to one when an atom's similarity exceeds $k_0$ (polymorph-like) and close to zero otherwise, so $s\approx 0$ in the parent phase and grows toward $N$ as the polymorph forms. It is this coordinate that the metadynamics bias acts on to drive the parent-to-polymorph transformation.

The WTMetaD bias is then applied along $s$. The history-dependent bias helps the system escape long-lived parent-phase-like or crystal-like basins and promotes exploration of the transition region. The time evolution of the ES-based crystallization coordinate in Fig.~\ref{fig:supp-metadynamics-es}C shows repeated fluctuations across the relevant CV range, indicating sufficient sampling for free-energy surface reconstruction.

\subparagraph*{2.2.3 Free-energy surface and nucleation driving force}

The reconstructed free-energy surface (FES) describes the relative free energy of parent-phase-like and crystal-like configurations along the ES coordinate. Because metadynamics biases the trajectory, the free energy is not read directly from the biased histogram. Instead, reweighting corrects for the deposited bias to recover an estimate of the unbiased probability distribution $P(s)$ of the crystallization coordinate $s$, from which the FES follows as
\begin{equation}
F(s)
=
-k_\mathrm{B}T\ln P(s),
\label{eq:fes-probability}
\end{equation}
\noindent\ 
where $k_\mathrm{B}$ is the Boltzmann constant and $T$ is the simulation temperature. Lower-probability regions correspond to higher free energy, and the relative statistical weights of the parent-phase-like and polymorph-like basins in the reweighted distribution determine the parent-to-polymorph free-energy difference. A representative FES is shown in Fig.~\ref{fig:supp-metadynamics-es}D, together with the basin assignment used to estimate this difference.

To convert the reconstructed FES into a driving force, the continuous ES coordinate is divided into parent-phase-like and polymorph-like regions. The statistical weight of each region is then estimated by integrating the reconstructed probability distribution over the corresponding CV range:
\begin{equation}
Z_\ell
\propto
\int_{\Omega_\ell} P(s)\,ds,
\qquad
Z_i
\propto
\int_{\Omega_i} P(s)\,ds,
\label{eq:basin-weights}
\end{equation}
\noindent\ 
where $\Omega_\ell$ and $\Omega_i$ denote the parent-phase-like (liquid-like) and polymorph-like regions of the ES coordinate, respectively; throughout, the subscript $\ell$ refers to the parent phase and $i$ to the crystalline polymorph. Comparing these basin weights gives the parent-to-polymorph free-energy difference:
\begin{equation}
\Delta F_{\ell\to i}
=
-k_\mathrm{B}T
\ln\left(
\frac{Z_i}{Z_\ell}
\right),
\label{eq:basin-free-energy}
\end{equation}
\noindent\ 
where $Z_\ell$ and $Z_i$ are the basin-restricted statistical weights. Below the melting temperature, the polymorph basin lies lower, so $\Delta F_{\ell\to i}<0$.

\begin{figure}
  \centering
  \includegraphics[width=0.9\textwidth]{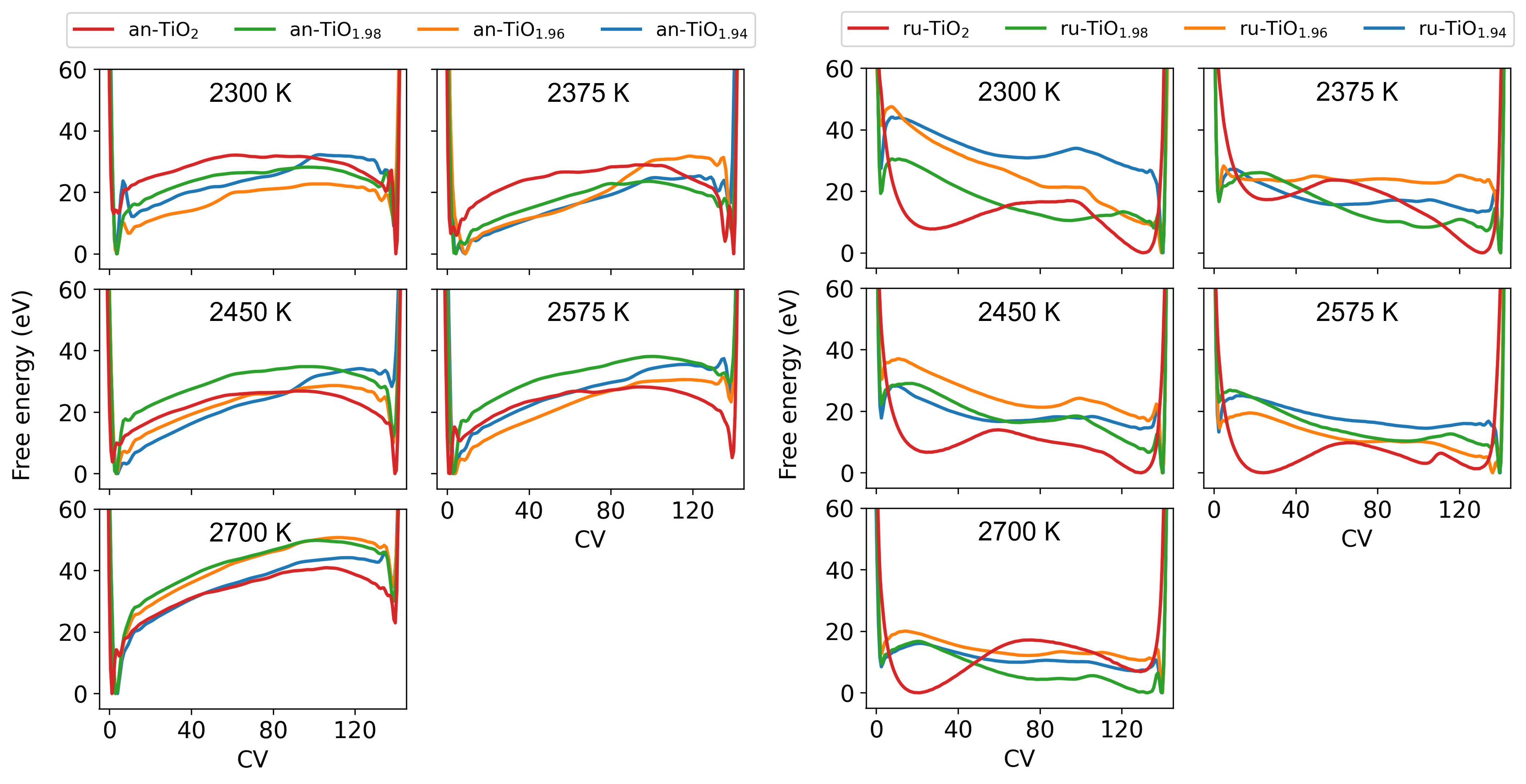}
  \caption{\small\textbf{Reconstructed free-energy surfaces across temperature and stoichiometry.} Free-energy surfaces reconstructed along the ES coordinate for the parent-to-anatase and parent-to-rutile transformations. The left and right blocks correspond to anatase and rutile references, respectively. Each subplot corresponds to one temperature. Line colors denote oxygen stoichiometry: red for TiO$_2$, green for TiO$_{1.98}$, orange for TiO$_{1.96}$, and blue for TiO$_{1.94}$.}
  \label{fig:supp-free-energy-surfaces}
\end{figure}

\begin{figure}
  \centering
  \includegraphics[width=0.9\textwidth]{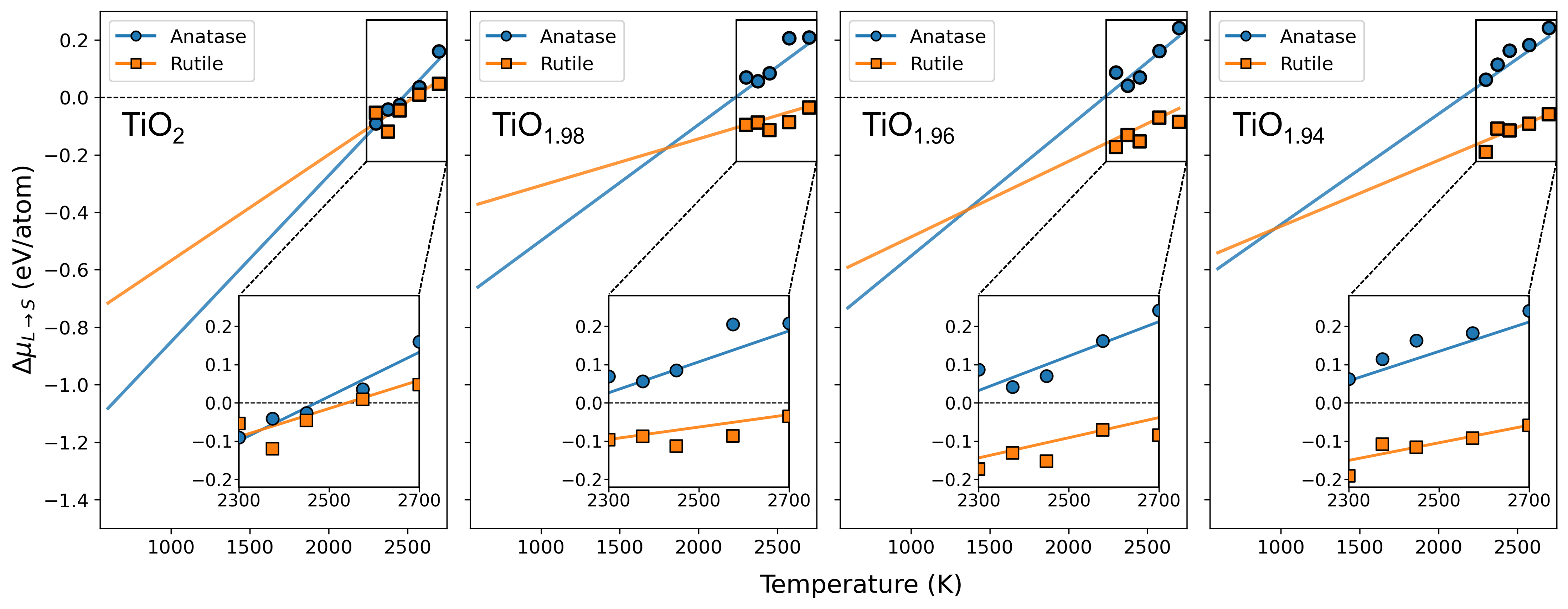}
  \caption{\small\textbf{Driving force for anatase and rutile nucleation.} $|\Delta\mu_{\ell\to i}|$ is plotted as a function of temperature for anatase and rutile. Subplots are organized by oxygen stoichiometry from TiO$_2$ to TiO$_{1.94}$ from left to right. Blue circles and blue lines denote anatase metadynamics-derived values and fits, respectively; orange squares and orange lines denote the corresponding rutile values and fits. Insets enlarge the calculated data points.}
  \label{fig:supp-driving-force}
\end{figure}

For each polymorph, temperature, and stoichiometry, the resulting $\Delta F_{\ell\to i}$ is normalized by the number of Ti-centered environments:
\begin{equation}
\left|\Delta\mu_{\ell\to i}\right|
=
\frac{\left|\Delta F_{\ell\to i}\right|}{N_\mathrm{CV}}.
\label{eq:driving-force-per-environment}
\end{equation}
\noindent\ 
This normalization assumes $\left|\Delta F_{\ell\to i}\right|\approx N_\mathrm{CV}\left|\Delta\mu_{\ell\to i}\right|$, where $N_\mathrm{CV}$ is the number of Ti atoms in the simulation box and equals the $N$ summed in the collective variable $s$ of Section 2.2.2. The numerical metadynamics and CV parameters (Gaussian height and width, deposition stride, $\sigma$, $k_0$, $p$, $q$, and the basin boundaries $\Omega_\ell$, $\Omega_i$) are listed in Methods. The same basin-weight analysis is repeated for every polymorph, temperature, and oxygen stoichiometry. Figure~\ref{fig:supp-free-energy-surfaces} summarizes the reconstructed FESs for anatase and rutile. Each subplot corresponds to one temperature, and the curves compare oxygen stoichiometries from TiO$_2$ to TiO$_{1.94}$. The resulting values are plotted as a function of temperature in Fig.~\ref{fig:supp-driving-force}. To summarize the temperature dependence, the data are fit using a Turnbull-type relation:
\begin{equation}
\left|\Delta\mu_{\ell\to i}\right|(T)
=
\Delta H_{m,i}
\left(1-\frac{T}{T_{m,i}}\right).
\label{eq:turnbull-driving-force}
\end{equation}
\noindent\ 
Here, $\Delta H_{m,i}$ is the melting enthalpy and $T_{m,i}$ is the fitted melting temperature for polymorph $i$. The zero crossing of each fitted line gives $T_m$. For stoichiometric TiO$_2$, the fitted melting temperatures are approximately 2450~K for the parent-to-anatase transformation and 2550~K for the parent-to-rutile transformation, close to the melting temperature of the employed potential (2400~K)~\cite{zhang2023investigating}. This agreement supports the consistency of the reconstructed FESs and the Turnbull-type extrapolation. In Fig.~\ref{fig:supp-driving-force}, the scatter points represent metadynamics-derived values, the solid lines represent the Turnbull-type fits, and the subplots compare different oxygen stoichiometries.

\paragraph*{2.3 Seeding method for $n_{c,i}$}

After obtaining $|\Delta\mu_{\ell\to i}|$, the second required quantity is the critical nucleus size $n_{c,i}$. Rather than inferring it from a separately determined interfacial energy, $n_{c,i}$ is measured directly from unbiased seeding simulations. In a seeding simulation, a preformed crystalline cluster of a chosen size is embedded in the supercooled parent phase and evolved under unbiased dynamics. Clusters smaller than the critical size tend to dissolve, whereas larger clusters tend to grow; scanning the seed size therefore identifies the size at which growth and dissolution are equally probable. This crossover size is the critical nucleus size $n_{c,i}$ that enters the barrier expression of Section 2.1, expressed here as the number of atoms.

Spherical anatase and rutile seeds spanning a range of sizes are inserted at the center of a supercooled parent-phase TiO$_{2-x}$ supercell, with the surrounding region melted to establish the supercooled liquid environment around each seed. Each seeded configuration is equilibrated and then propagated under unbiased molecular dynamics at each seed size, following the seeding protocol described in Methods. Throughout each trajectory, the seed is classified as growing or dissolving based on its normalized crystallinity along the ES coordinate introduced in Section 2.2.2, defined as the fraction of seed-region Ti environments that remain polymorph-like. A threshold of 0.8 separates the two outcomes. This threshold was set above the range of thermal fluctuations so that thermal noise alone would not change the growth/dissolution assignment. The critical size $n_{c,i}$ is identified as the seed size at which the growth and dissolution probabilities are equal.

\begin{figure}
  \centering
  \includegraphics[width=0.9\textwidth]{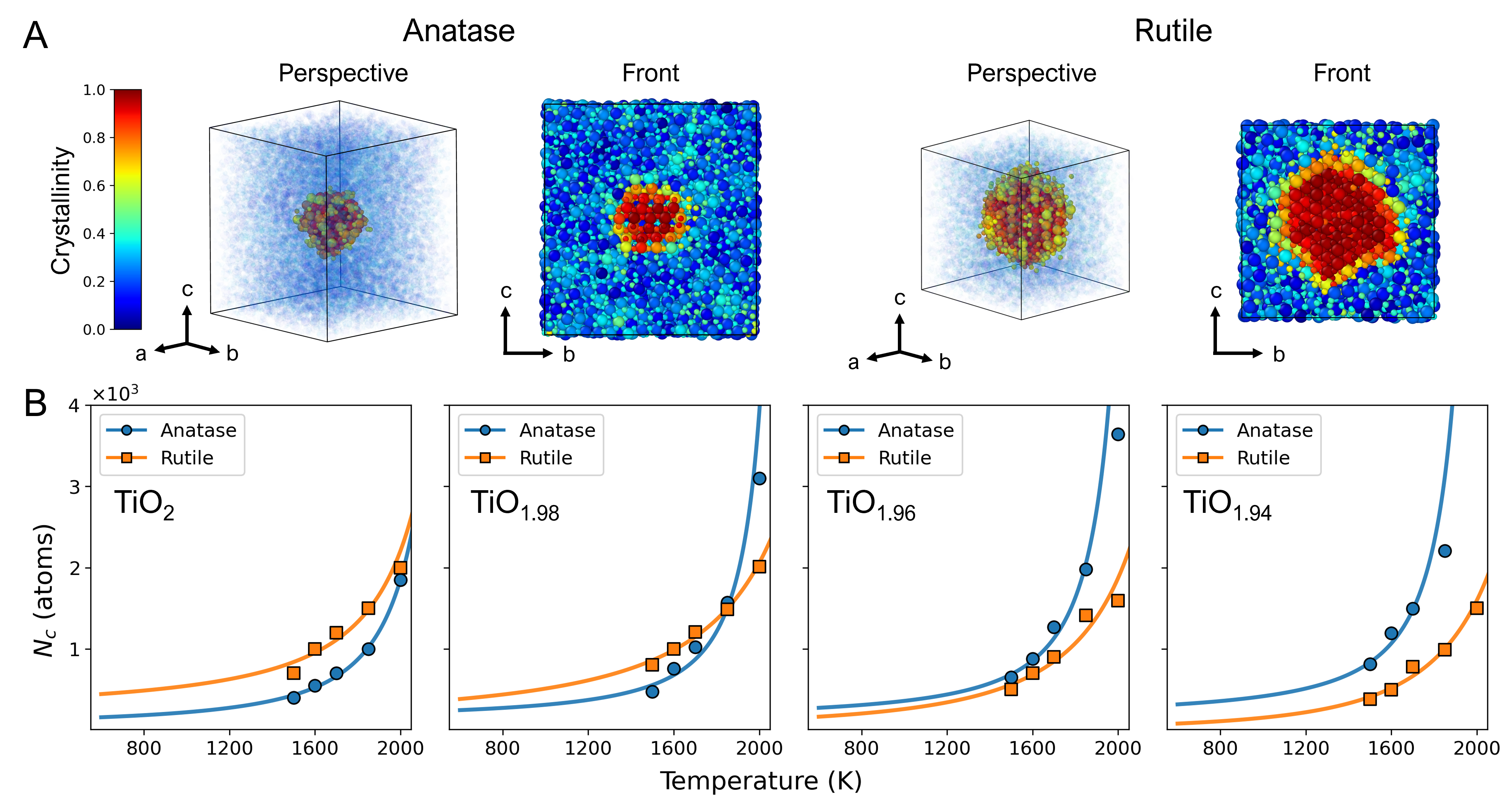}
  \caption{\small\textbf{Seeding method and critical nucleus size.} \textbf{(A)} Representative anatase and rutile seed configurations shown in perspective and front views. Color denotes crystallinity. \textbf{(B)} Critical nucleus size $n_c$ as a function of temperature and oxygen stoichiometry. Blue circles and curves denote anatase, and orange squares and curves denote rutile. Points represent seeding results; curves show the reconstructed critical-size trends.}
  \label{fig:supp-seeding-critical-size}
\end{figure}

Fig.~\ref{fig:supp-seeding-critical-size} summarizes the seeding analysis. Fig.~\ref{fig:supp-seeding-critical-size}A shows representative anatase and rutile seed configurations in perspective and front views, colored by crystallinity to distinguish the ordered seed from the surrounding supercooled liquid. Fig.~\ref{fig:supp-seeding-critical-size}B reports the resulting critical nucleus sizes as functions of oxygen stoichiometry and temperature. Each seeding-derived $n_{c,i}$ value was combined with the corresponding $|\Delta\mu_{\ell\to i}|$ and the critical-size relation in Section 2.1 to back-calculate an effective nucleus--parent interfacial energy $\gamma_i$ for that condition:
\begin{equation}
\gamma_i
=
\left[
\frac{
3\rho_{s,i}^{2}
\left|\Delta\mu_{\ell\to i}\right|^{3}
n_{c,i}
}
{32\pi}
\right]^{1/3},
\label{eq:effective-interfacial-energy}
\end{equation}
\noindent\ 
where $\rho_{s,i}$ is the atomic density of polymorph $i$. Finally, the smooth curves in Fig.~\ref{fig:supp-seeding-critical-size}B were reconstructed by inserting the fitted driving force $|\Delta\mu_{\ell\to i}(T)|$ and fitted temperature-dependent effective interfacial energy $\gamma_i(T)$ back into the critical-size relation:
\begin{equation}
n_{c,i}(T)
=
\frac{
32\pi\gamma_i(T)^3
}
{
3\rho_{s,i}^{2}
\left|\Delta\mu_{\ell\to i}(T)\right|^{3}
}.
\label{eq:critical-size-reconstruction}
\end{equation}
\noindent\ 
Thus, the solid curves in Fig.~\ref{fig:supp-seeding-critical-size}B are reconstructed from the seeding-derived $n_{c,i}$ values, enhanced-sampling driving forces, and effective interfacial energies, rather than fitted directly to the $n_{c,i}$ points.

\paragraph*{2.4 Barrier calculation}

We now return to the barrier expression introduced in Section 2.1. The enhanced-sampling analysis in Section 2.2 provides the driving force $|\Delta\mu_{\ell\to i}|$, while the seeding analysis in Section 2.3 provides the corresponding critical nucleus size $n_{c,i}$. The continuous temperature-dependent forms of these quantities, given by Eqs.~\eqref{eq:turnbull-driving-force} and \eqref{eq:critical-size-reconstruction}, are evaluated on a common temperature axis and combined to obtain the nucleation barrier for each polymorph:
\begin{equation}
\Delta G_i^*
\simeq
\frac{1}{2}
\left|\Delta\mu_{\ell\to i}\right|n_{c,i}.
\label{eq:barrier-assembly}
\end{equation}
\noindent\ 
The barrier difference used in the main manuscript is
\begin{equation}
\Delta G^*_{\mathrm{rutile}}
-
\Delta G^*_{\mathrm{anatase}}.
\label{eq:barrier-difference}
\end{equation}
\noindent\ 
Positive values indicate a lower anatase barrier, whereas negative values indicate a lower rutile barrier. Evaluating this barrier difference across temperature and oxygen stoichiometry yields the barrier map shown in main-text Fig. 3A.

\subsubsection*{Supplementary Note 3 -- Coupled-flux model}

This note describes the coupled-flux model (CFM) used to quantify atomic exchange between the cluster and parent phase through the intermediate shell. In the model, shell-mediated exchange is represented through the pathway-dependent rate factor $A_i$ in the nucleation-rate expression. The model follows the coupled-flux (linked-flux) treatment of Refs.~\cite{kelton2000time,kelton2010nucleation}. We restate it in the notation of this work and apply it to TiO$_{2-x}$.

\paragraph*{3.1 Motivation for a coupled-flux description}

\begin{figure}
  \centering
  \includegraphics[width=0.9\textwidth]{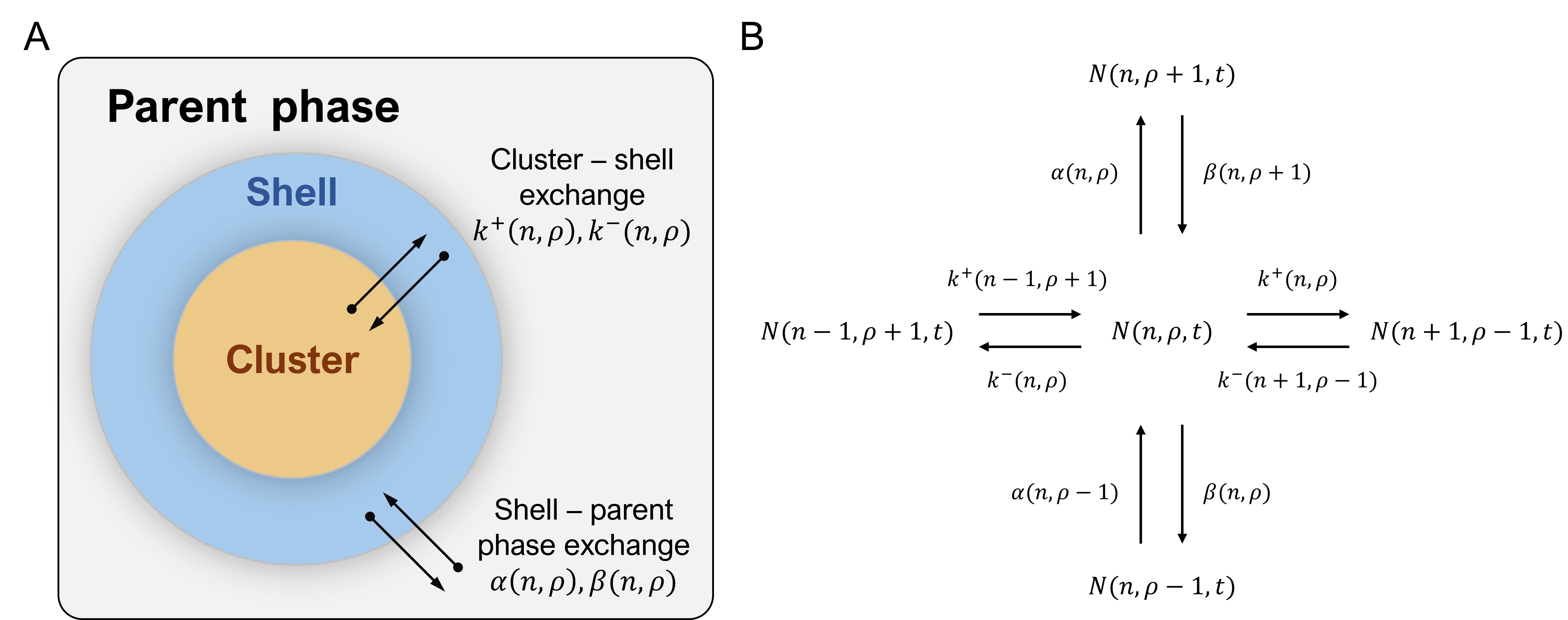}
  \caption{\small\textbf{Coupled-flux description of shell-mediated cluster growth.} \textbf{(A)} Physical picture of a crystal-like cluster surrounded by an intermediate shell and a parent phase. \textbf{(B)} Rate processes in $(n,\rho)$ space. Cluster--shell attachment and detachment change both $n$ and $\rho$ through $k^+$ and $k^-$, whereas diffusion into and out of the shell changes $\rho$ through $\alpha$ and $\beta$ at fixed $n$.}
  \label{fig:supp-coupled-flux-schematic}
\end{figure}

Figure~\ref{fig:supp-coupled-flux-schematic} summarizes the physical picture used in this note. In this description, the growing cluster is surrounded by an intermediate shell that mediates atomic exchange with the parent phase. The coupled-flux model distinguishes two exchange steps: diffusion between the parent phase and the shell, and attachment and detachment between the shell and the cluster. These steps are coupled through the shell population.

\paragraph*{3.2 Shell-mediated nucleation rate}

Main-text Fig. 4 shows that the intermediate shell is a structurally active region whose local environments evolve during nucleation. We therefore incorporate its contribution into the kinetic factor of the nucleation rate. The coupled-flux model describes the time evolution of the cluster population; in the steady-state limit, the nucleation rate for a polymorphic pathway $i$ is written as
\begin{equation}
J_i^\mathrm{CFM}
=
A_i^\mathrm{CFM}
\exp\left[
-\frac{\Delta G_i^*}{k_\mathrm{B}T}
\right].
\label{eq:cfm-target-rate-expression}
\end{equation}
\noindent\
The barrier term is the same as in Supplementary Note 2; the shell contribution enters through the kinetic factor:
\begin{equation}
A_i^\mathrm{CFM}
\simeq
\frac{
\alpha_i(n_{c,i},0)
}{
k_i^+(n_{c,i},1)
}
A_i^\mathrm{CNT}.
\label{eq:cfm-target-rate-factor}
\end{equation}
\noindent\
Here, $\Delta G_i^*$ is the pathway-dependent critical nucleation barrier, $n_{c,i}$ is the critical nucleus size, and $A_i^\mathrm{CNT}$ is the CNT kinetic prefactor for the same pathway. The ratio $\alpha_i(n_{c,i},0)/k_i^+(n_{c,i},1)$ represents the shell-mediated contribution: the numerator describes diffusion into an empty shell around the critical cluster from the parent phase, whereas the denominator describes attachment from a one-atom shell into that cluster. The rest of this note defines the two variables used to describe this process, the cluster size $n$ and shell population $\rho$, and then derives the rates $\alpha_i$ and $k_i^+$ that appear in Eq.~\eqref{eq:cfm-target-rate-factor}.

\paragraph*{3.3 Cluster distribution including shell population}

To include shell-mediated exchange, the cluster distribution is resolved in terms of the cluster size $n$ and shell population $\rho$. Following the three-region description of the coupled-flux model, we distinguish the crystal-like cluster, its surrounding shell, and the remaining parent phase.

We describe the state of a cluster by two variables,
\begin{equation}
(n,\rho),
\label{eq:cfm-state-variables}
\end{equation}
\noindent\ 
where $n$ is the number of atoms in the cluster, and $\rho$ is the shell-population variable, defined as the number of atoms in the nearest-neighbor shell around that cluster. For each cluster size, the shell has a finite number of available sites, denoted by $\rho_{\max}(n)$. This capacity increases with cluster size because the number of shell sites scales with the cluster surface area.

Let $N_i(n,\rho,t)$ denote the population of pathway-$i$ clusters with size $n$ and shell population $\rho$ at time $t$. The coupled-flux model follows transitions between these states through two coupled processes. Attachment and detachment transfer atoms between the shell and the cluster, changing both $n$ and $\rho$, whereas diffusion transfers atoms between the parent phase and the shell, changing $\rho$ at fixed $n$.

\paragraph*{3.4 Equilibrium distribution in $(n,\rho)$ space}

The rate constants below are constrained using an equilibrium cluster distribution written in the same variables $(n,\rho)$. For pathway $i$, we write this distribution as
\begin{equation}
N_i^\mathrm{eq}(n,\rho)
=
N_\mathrm{sol}
\exp\left[-\frac{W_i(n)}{k_\mathrm{B}T}\right]
P_i(\rho\mid n),
\label{eq:cfm-equilibrium-distribution}
\end{equation}
\noindent\ 
where $W_i(n)$ is the work of forming a pathway-$i$ cluster of size $n$, and $P_i(\rho\mid n)$ is the normalized probability of having $\rho$ atoms in the nearest-neighbor shell around that cluster. The factor $N_\mathrm{sol}$ denotes the number of solute atoms in the parent phase.

The shell probability follows from the configurational entropy of distributing atoms between the shell and the parent phase. The $\rho$ atoms occupy the $\rho_{\max}(n)$ sites available in the shell around a cluster of size $n$, and are drawn from a parent phase of $N_\mathrm{sol}$ atoms distributed over $N_\mathrm{sites}$ sites, so that $N_\mathrm{sol}/N_\mathrm{sites}$ is the parent-phase atom fraction. Combining the shell and parent occupancies gives
\begin{equation}
P_i(\rho\mid n)
\propto
\underbrace{
\frac{
\rho_{\max}(n)!
}{
\rho!\,
\left[\rho_{\max}(n)-\rho\right]!
}
}_{\text{shell occupancy}}
\;
\underbrace{
\frac{
N_\mathrm{sol}!\,
\left(N_\mathrm{sites}-N_\mathrm{sol}\right)!
}{
\left(N_\mathrm{sol}-\rho\right)!\,
\left(N_\mathrm{sites}-N_\mathrm{sol}+\rho\right)!
}
}_{\text{parent depletion}}.
\label{eq:cfm-shell-probability}
\end{equation}
\noindent\ 
For the cluster sizes relevant to nucleation, we assume that $N_\mathrm{sol}\gg\rho$. If the initial parent phase is also dilute, Eq.~\eqref{eq:cfm-shell-probability} can be simplified using Stirling's approximation as
\begin{equation}
P_i(\rho\mid n)
=
\varepsilon_i(n)
\frac{
\rho_{\max}(n)!
}{
\rho!\,
\left[\rho_{\max}(n)-\rho\right]!
}
\left(
\frac{N_\mathrm{sol}}
{N_\mathrm{sites}-N_\mathrm{sol}}
\right)^\rho,
\label{eq:cfm-shell-probability-dilute}
\end{equation}
\noindent\ 
where $\varepsilon_i(n)$ is the normalization constant chosen so that $\sum_{\rho=0}^{\rho_{\max}(n)}P_i(\rho\mid n)=1$. This equilibrium distribution is used below to impose detailed balance on paired rate constants.

\paragraph*{3.5 Rate processes in $(n,\rho)$ space}

The coupled-flux model treats cluster growth and shell exchange as one-atom events (Fig.~\ref{fig:supp-coupled-flux-schematic}B). Two rates describe exchange between the cluster and the shell. The attachment rate $k_i^+(n,\rho)$ incorporates one shell atom into the cluster, changing the state from $(n,\rho)$ to $(n+1,\rho-1)$. The detachment rate $k_i^-(n,\rho)$ transfers one atom from the cluster back to the shell, changing the state from $(n,\rho)$ to $(n-1,\rho+1)$.

Two additional rates describe exchange between the shell and the parent phase. The rate $\alpha_i(n,\rho)$ adds one atom to the shell, changing the state from $(n,\rho)$ to $(n,\rho+1)$, whereas $\beta_i(n,\rho)$ removes one atom from the shell. Together, these four rates distinguish diffusion between the parent phase and the shell from attachment and detachment at the cluster.

\paragraph*{3.6 Detailed-balance expressions}

At equilibrium, each exchange process must be balanced by its opposite process in $(n,\rho)$ space. For diffusion between the parent phase and the shell, a cluster with shell population $\rho-1$ gains one atom when atoms diffuse into the shell, whereas a cluster with shell population $\rho$ loses one atom when atoms diffuse out of the shell:
\begin{equation}
\alpha_i(n,\rho-1)
N_i^\mathrm{eq}(n,\rho-1)
=
\beta_i(n,\rho)
N_i^\mathrm{eq}(n,\rho).
\label{eq:cfm-detailed-balance-shell-parent}
\end{equation}
\noindent\
The rate for diffusion out of the shell is proportional to the number of atoms in the shell and to the parent-phase mobility. The detailed-balance condition in Eq.~\eqref{eq:cfm-detailed-balance-shell-parent} fixes only the ratio $\alpha_i/\beta_i$; assuming the forward and reverse rates share this equilibrium factor symmetrically, and using a pathway-dependent jump distance $\lambda_i$, a parent-phase diffusion coefficient $D_{\mathrm{parent}}$, and a scaling factor $\xi_i$, the paired rates are written as
\begin{equation}
\begin{aligned}
\alpha_i(n,\rho-1)
&=
\xi_i\rho
\frac{D_{\mathrm{parent}}}{\lambda_i^2}
\left[
\frac{\rho_{\max}(n)-\rho+1}{\rho}
\right]^{1/2}
\left(
\frac{N_\mathrm{sol}}
{N_\mathrm{sites}-N_\mathrm{sol}}
\right)^{1/2},\\
\beta_i(n,\rho)
&=
\xi_i\rho
\frac{D_{\mathrm{parent}}}{\lambda_i^2}
\left[
\frac{\rho_{\max}(n)-\rho+1}{\rho}
\right]^{-1/2}
\left(
\frac{N_\mathrm{sol}}
{N_\mathrm{sites}-N_\mathrm{sol}}
\right)^{-1/2}.
\end{aligned}
\label{eq:cfm-diffusion-rates}
\end{equation}
\noindent\
The scaling factor $\xi_i$ accounts for the fact that an atom that diffuses out of the shell does not immediately become equivalent to an atom in the fully mixed parent phase. Matching the model's large-cluster growth to the known diffusion-limited growth rate fixes its value:
\begin{equation}
\xi_i
=
\frac{(4\pi)^{2/3}}{2}
\left(3\bar{v}_i\right)^{1/3}
\lambda_i^2
\left(N_\mathrm{sites}X_{\infty}\right)^{1/2},
\label{eq:cfm-xi-scaling-factor}
\end{equation}
\noindent\
where $X_{\infty}=N_\mathrm{sol}/N_\mathrm{sites}$ is the fraction of atoms in the parent phase far from the growing cluster, and $\bar{v}_i$ is the atomic volume of polymorph $i$.

For exchange between the shell and the cluster, detailed balance between attachment and detachment requires
\begin{equation}
k_i^+(n,\rho)
N_i^\mathrm{eq}(n,\rho)
=
k_i^-(n+1,\rho-1)
N_i^\mathrm{eq}(n+1,\rho-1).
\label{eq:cfm-detailed-balance-cluster-shell}
\end{equation}
\noindent\
Attachment is controlled by diffusion within the shell adjacent to a pathway-$i$ cluster, represented by $6D_{\mathrm{shell},i}/\lambda_i^2$, where $D_{\mathrm{shell},i}$ is the effective diffusion coefficient within that shell. With $\delta W_i(n)=W_i(n+1)-W_i(n)$, and again splitting the equilibrium factor symmetrically between the forward and reverse rates, applying Eq.~\eqref{eq:cfm-detailed-balance-cluster-shell} with the equilibrium distribution above gives
\begin{equation}
\begin{aligned}
k_i^+(n,\rho)
&=
\rho
\frac{6D_{\mathrm{shell},i}}{\lambda_i^2}
\exp\left[
-\frac{\delta W_i(n)}{2k_\mathrm{B}T}
\right]
G_i(n,\rho),\\
k_i^-(n+1,\rho-1)
&=
\rho
\frac{6D_{\mathrm{shell},i}}{\lambda_i^2}
\exp\left[
+\frac{\delta W_i(n)}{2k_\mathrm{B}T}
\right]
\frac{1}{G_i(n,\rho)}.
\end{aligned}
\label{eq:cfm-attachment-rates}
\end{equation}
\noindent\
The factor $G_i(n,\rho)$ accounts for the change in the number of possible shell and parent-phase configurations when one atom attaches to the cluster:
\begin{equation}
G_i(n,\rho)
=
\left[
\frac{\varepsilon_i(n)}{\varepsilon_i(n+1)}
\frac{\rho_{\max}(n)!}{\rho_{\max}(n+1)!}
\frac{1}{\rho}
\frac{\left[\rho_{\max}(n+1)-\rho+1\right]!}
{\left[\rho_{\max}(n+1)-\rho\right]!}
\left(
\frac{N_\mathrm{sol}}
{N_\mathrm{sites}-N_\mathrm{sol}}
\right)
\right]^{-1/2}.
\label{eq:cfm-configurational-correction}
\end{equation}
\noindent\
Here, $D_{\mathrm{parent}}$ controls diffusion between the parent phase and the shell, whereas $D_{\mathrm{shell},i}$ controls mobility within the shell adjacent to a pathway-$i$ cluster. Thus, the diffusion pair and the attachment--detachment pair are both tied to the same equilibrium distribution before the rate factor is reduced near the critical region.

\paragraph*{3.7 Steady-state reduction to the rate factor}

At steady state, the nucleation flux is independent of time and can be evaluated at the critical state. For pathway $i$, this critical state is specified by the critical nucleus size $n_{c,i}$. We therefore evaluate the rate expressions from Sections 3.5 and 3.6 at $n_{c,i}$ to obtain the rate factor multiplying the barrier term in the nucleation-rate expression.

The detailed-balance relations in Section 3.6 express the reverse rates in terms of the corresponding forward rates and the equilibrium distribution:
\begin{equation}
\begin{aligned}
\beta_i(n,\rho)
&=
\alpha_i(n,\rho-1)
\frac{
N_i^\mathrm{eq}(n,\rho-1)
}{
N_i^\mathrm{eq}(n,\rho)
},\\
k_i^-(n+1,\rho-1)
&=
k_i^+(n,\rho)
\frac{
N_i^\mathrm{eq}(n,\rho)
}{
N_i^\mathrm{eq}(n+1,\rho-1)
}.
\end{aligned}
\label{eq:cfm-reverse-rates-from-balance}
\end{equation}
\noindent\
In the dilute-shell limit near the critical state, shell-mediated exchange is represented as a serial sequence: an atom first diffuses into the empty shell around the critical cluster and then attaches to the cluster. These two forward steps are represented by $\alpha_i(n_{c,i},0)$ and $k_i^+(n_{c,i},1)$, respectively. Because the steps act in series, the slower one limits the combined process. In the diffusion-limited regime relevant here, diffusion into the shell is slower than attachment, so the coupled rate is reduced relative to the CNT rate by the ratio $\alpha_i(n_{c,i},0)/k_i^+(n_{c,i},1)$. The coupled steady-state rate is therefore written in terms of the CNT rate as
\begin{equation}
J_i^\mathrm{CFM}
\simeq
\frac{
\alpha_i(n_{c,i},0)
}{
k_i^+(n_{c,i},1)
}
J_i^\mathrm{CNT}.
\label{eq:cfm-steady-state-rate-scaling}
\end{equation}
\noindent\
Here, $J_i^\mathrm{CNT}$ denotes the CNT rate obtained when the critical-state kinetics are represented by a single attachment step. Written in the activated form used in the main text,
\begin{equation}
J_i
=
A_i
\exp\left[
-\frac{\Delta G_i^*}{k_\mathrm{B}T}
\right],
\label{eq:cfm-steady-state-rate-form}
\end{equation}
\noindent\
the barrier term $\Delta G_i^*$ is the critical nucleation barrier, while the rate factor $A_i$ contains the kinetic contribution from the coupled shell process. Because the exponential barrier term is kept in the same form, Eq.~\eqref{eq:cfm-steady-state-rate-scaling} gives the corresponding rate-factor scaling
\begin{equation}
A_i^\mathrm{CFM}
\simeq
\frac{
\alpha_i(n_{c,i},0)
}{
k_i^+(n_{c,i},1)
}
A_i^\mathrm{CNT}.
\label{eq:cfm-rate-factor-scaling}
\end{equation}

Using the rate expressions in Eqs.~\eqref{eq:cfm-diffusion-rates} and \eqref{eq:cfm-attachment-rates}, the ratio in Eq.~\eqref{eq:cfm-rate-factor-scaling} is
\begin{equation}
\frac{
\alpha_i(n_{c,i},0)
}{
k_i^+(n_{c,i},1)
}
=
\frac{
\xi_i D_{\mathrm{parent}}
}{
6D_{\mathrm{shell},i}
}
\left[
\rho_{\max}(n_{c,i})
\right]^{1/2}
\left(
\frac{N_\mathrm{sol}}
{N_\mathrm{sites}-N_\mathrm{sol}}
\right)^{1/2}
\exp\left[
\frac{\delta W_i(n_{c,i})}{2k_\mathrm{B}T}
\right]
\frac{1}{G_i(n_{c,i},1)}.
\label{eq:cfm-critical-rate-ratio}
\end{equation}
\noindent\
For the TiO$_{2-x}$ systems, the critical sizes obtained from the seeding analysis contain hundreds to thousands of atoms (Fig.~\ref{fig:supp-seeding-critical-size}B). The addition of a single atom therefore produces only a small change in the cluster free energy at the top of the barrier, so we take $\delta W_i(n_{c,i})\ll k_\mathrm{B}T$ and set the exponential factor in Eq.~\eqref{eq:cfm-critical-rate-ratio} to unity. The same large-cluster limit also simplifies the slowly varying configurational factors in $G_i(n_{c,i},1)$. Specifically, we take
\begin{equation}
\frac{\varepsilon_i(n_{c,i})}{\varepsilon_i(n_{c,i}+1)}
\simeq
1,
\qquad
\frac{\rho_{\max}(n_{c,i})!}{\rho_{\max}(n_{c,i}+1)!}
\frac{\rho_{\max}(n_{c,i}+1)!}
{\left[\rho_{\max}(n_{c,i}+1)-1\right]!}
\simeq
1,
\label{eq:cfm-large-cluster-approximations}
\end{equation}
\noindent\
where the first approximation states that the shell-population normalization changes weakly when $n_{c,i}$ increases by one atom, and the second states that the shell capacity changes weakly on the same step. With $\rho=1$, these approximations give
\begin{equation}
G_i(n_{c,i},1)
\simeq
\left(
\frac{N_\mathrm{sol}}
{N_\mathrm{sites}-N_\mathrm{sol}}
\right)^{-1/2}.
\label{eq:cfm-g-critical-approximation}
\end{equation}
\noindent\
Substituting Eq.~\eqref{eq:cfm-g-critical-approximation} into Eq.~\eqref{eq:cfm-critical-rate-ratio} gives
\begin{equation}
\frac{
\alpha_i(n_{c,i},0)
}{
k_i^+(n_{c,i},1)
}
\simeq
\frac{
\xi_i D_{\mathrm{parent}}
}{
6D_{\mathrm{shell},i}
}
\left[
\rho_{\max}(n_{c,i})
\right]^{1/2}
\left(
\frac{N_\mathrm{sol}}
{N_\mathrm{sites}-N_\mathrm{sol}}
\right).
\label{eq:cfm-reduced-critical-rate-ratio}
\end{equation}
\noindent\
The parameters entering Eq.~\eqref{eq:cfm-reduced-critical-rate-ratio} are summarized in Table~\ref{tab:cfm-rate-ratio-inputs}.

\begin{table}[ht]
\centering
\caption{\textbf{Parameters used to evaluate the shell-mediated contribution.} Description and source of each quantity in Eq.~\eqref{eq:cfm-reduced-critical-rate-ratio} for the TiO$_{2-x}$ evaluation.}
\label{tab:cfm-rate-ratio-inputs}
\begin{tabular}{@{}p{0.18\textwidth}p{0.35\textwidth}p{0.37\textwidth}@{}}
\hline
Parameter & Description & Source \\
\hline
$D_{\mathrm{parent}}$ & Parent-phase diffusion coefficient & Seeding simulations \\
$D_{\mathrm{shell},i}$ & Shell diffusion coefficient & Seeding simulations (atoms in the Fig.~4A shell) \\
$\rho_{\max}(n_{c,i})$ & Shell capacity at the critical size & $\rho_{\max}\simeq 4(n_{c,i})^{2/3}$~\cite{kelton2010nucleation}; $n_{c,i}$ from seeding (Fig.~S5B) \\
$N_\mathrm{sol}$ & Solute atoms in the parent phase & Seeding simulations \\
$N_\mathrm{sites}$ & Available sites (with $N_\mathrm{sol}$: composition factor $N_\mathrm{sol}/(N_\mathrm{sites}-N_\mathrm{sol})$) & Seeding simulations \\
$\lambda_i$ & Jump distance & Polymorph lattice parameter \\
$\bar{v}_i$ & Atomic volume & Polymorph crystal structure \\
$\xi_i$ & Shell--parent exchange scaling factor & Eq.~\eqref{eq:cfm-xi-scaling-factor} \\
\hline
\end{tabular}
\end{table}

\paragraph*{3.8 Total nucleation-rate expression}

As previewed in Section~3.2 (Eq.~\eqref{eq:cfm-target-rate-expression}), combining the barrier term with the shell-mediated rate factor gives
\begin{equation}
J_i^\mathrm{CFM}
=
A_i^\mathrm{CFM}
\exp\left[
-\frac{\Delta G_i^*}{k_\mathrm{B}T}
\right],
\label{eq:cfm-total-rate-expression}
\end{equation}
\noindent\
with
\begin{equation}
A_i^\mathrm{CFM}
\simeq
\frac{
\alpha_i(n_{c,i},0)
}{
k_i^+(n_{c,i},1)
}
A_i^\mathrm{CNT}
=
\frac{
\alpha_i(n_{c,i},0)
}{
k_i^+(n_{c,i},1)
}
N_{0,i} Z_i f_i^+.
\label{eq:cfm-total-rate-factor}
\end{equation}
\noindent\
The ratio $\alpha_i(n_{c,i},0)/k_i^+(n_{c,i},1)$ is the shell-mediated contribution derived in Eq.~\eqref{eq:cfm-reduced-critical-rate-ratio}. The remaining factor, $A_i^\mathrm{CNT}=N_{0,i}Z_if_i^+$, is the CNT kinetic prefactor. Here, $N_{0,i}$ is the number density of available nucleation sites, $Z_i=\left(\left|\Delta\mu_{\ell\to i}\right|/(6\pi k_\mathrm{B}T n_{c,i})\right)^{1/2}$ is the Zeldovich factor, which accounts for the spread of cluster sizes near the top of the barrier, and $f_i^+$ is the attachment frequency in the CNT expression.

Finally, the relative anatase--rutile nucleation preference is written as
\begin{equation}
\ln\frac{J_\mathrm{An}^\mathrm{CFM}}
{J_\mathrm{Ru}^\mathrm{CFM}}
=
-\frac{\Delta G_\mathrm{An}^*-\Delta G_\mathrm{Ru}^*}
{k_\mathrm{B}T}
+
\ln\frac{A_\mathrm{An}^\mathrm{CFM}}
{A_\mathrm{Ru}^\mathrm{CFM}}.
\label{eq:cfm-relative-rate-expression}
\end{equation}
\noindent\
Positive values indicate anatase-favored nucleation, whereas negative values indicate rutile-favored nucleation, as discussed in the main text.

\subsubsection*{Supplementary Note 4 -- Phase-field model}

Supplementary Note 4 describes the phase-field model used to connect composition-dependent nucleation preferences to growth on spatially heterogeneous composition fields, as shown in main-text Fig. 6. The calculation examines how composition-dependent seed identities affect the resulting domain patterns. The CFM assigns each prescribed seed an anatase or rutile identity based on its local composition, after which the phase-field model evolves the corresponding domains.

Main-text Fig. 6 presents the TiO$_2$ case, whereas Figs. S7--S9 show the corresponding results for TiO$_{1.98}$, TiO$_{1.96}$, and TiO$_{1.94}$, respectively. The supplementary cases apply the same model across the stoichiometry series and qualitatively illustrate how local composition can bias seed identity and subsequent polymorph growth.

Section 4.1 defines the phase fields and the continuum composition field. Section 4.2 explains how the coupled-flux model is used to assign the identities of prescribed phase-field seeds. Section 4.3 presents the phase-field free-energy functional used in the heterogeneity calculations. Section 4.4 summarizes the evolution equations and numerical constraints.

\paragraph*{4.1 Phase-field variables and composition field}

The phase-field model represents the two competing crystalline identities using two nonconserved order parameters,
\begin{equation}
\eta_\mathrm{A}(\mathbf{x},t),
\qquad
\eta_\mathrm{R}(\mathbf{x},t),
\label{eq:pfm-order-parameters}
\end{equation}
\noindent\ 
where $\eta_\mathrm{A}$ and $\eta_\mathrm{R}$ are coarse-grained order parameters for anatase-like and rutile-like material, respectively, at position $\mathbf{x}$ and time $t$. We refer to them below as the anatase and rutile phase fields. For either field, values approaching one indicate locally transformed material with the corresponding polymorph identity, whereas the untransformed state corresponds to both $\eta_\mathrm{A}$ and $\eta_\mathrm{R}$ being near zero.

The total transformed fraction and the untransformed parent fraction are defined as
\begin{equation}
q(\mathbf{x},t)
=
\eta_\mathrm{A}(\mathbf{x},t)
+
\eta_\mathrm{R}(\mathbf{x},t),
\qquad
\phi_\mathrm{P}(\mathbf{x},t)
=
1-q(\mathbf{x},t).
\label{eq:pfm-transformed-parent-fractions}
\end{equation}
\noindent\ 
Thus, $q$ represents the local fraction transformed into either crystalline identity, while $\phi_\mathrm{P}$ represents the remaining parent fraction.

The conserved composition field is denoted by $c(\mathbf{x},t)$. Here, $c$ is a normalized continuum coordinate for local oxygen composition, with larger values corresponding to greater oxygen deficiency. It describes local oxygen stoichiometry without explicitly counting or tracking individual oxygen vacancies. The initial condition is written as
\begin{equation}
c(\mathbf{x},0)
=
c_0+\delta c(\mathbf{x}),
\label{eq:pfm-initial-composition}
\end{equation}
\noindent\ 
where $c_0$ sets the nominal TiO$_{2-x}$ composition and $\delta c(\mathbf{x})$ introduces the imposed spatial heterogeneity. Details of the numerical construction and parameter values of $\delta c(\mathbf{x})$ are provided in Materials and Methods.

\paragraph*{4.2 CFM-based seed-identity assignment}

The phase-field calculation begins from prescribed seed sites whose placement is described in Materials and Methods. The CFM then defines the probabilities for anatase and rutile identities from the relative nucleation preference at the local initial composition.

For a seed site $s$ at $\mathbf{x}_s$, the composition is taken from the initial field:
\begin{equation}
c_s
=
c(\mathbf{x}_s,0)
=
c_0+\delta c(\mathbf{x}_s).
\label{eq:pfm-seed-local-composition}
\end{equation}
\noindent\ 
To avoid extrapolating the CFM preference beyond the composition range covered by the CFM calculations, the nearest boundary value was used for any input outside this range. The CFM-derived local nucleation preference at temperature $T$ and seed composition $c_s$ is defined as
\begin{equation}
\lambda_s
=
\ln
\left[
\frac{
J_\mathrm{An}^\mathrm{CFM}(T,c_s)
}{
J_\mathrm{Ru}^\mathrm{CFM}(T,c_s)
}
\right],
\label{eq:pfm-seed-lambda}
\end{equation}
\noindent\ 
Here, $J_\mathrm{An}^\mathrm{CFM}$ and $J_\mathrm{Ru}^\mathrm{CFM}$ are the CFM nucleation rates for anatase and rutile, respectively. Positive $\lambda_s$ favors anatase, whereas negative $\lambda_s$ favors rutile.

Using the two nucleation rates as competing weights for seed identity, the corresponding probabilities are
\begin{equation}
P_\mathrm{A}(s)
=
\frac{1}{1+\exp(-\lambda_s)},
\qquad
P_\mathrm{R}(s)
=
1-P_\mathrm{A}(s).
\label{eq:pfm-seed-probability}
\end{equation}
\noindent\ 
The stochastic seed identity is specified by
\begin{equation}
\Pr(i_s=\mathrm{A}\mid s)=P_\mathrm{A}(s),
\qquad
\Pr(i_s=\mathrm{R}\mid s)=P_\mathrm{R}(s).
\label{eq:pfm-seed-identity-sampling}
\end{equation}
\noindent\ 
The sampled outcome $i_s=\mathrm{A}$ or $\mathrm{R}$ determines whether the seed profile is inserted into $\eta_\mathrm{A}$ or $\eta_\mathrm{R}$, respectively. The seed geometry is prescribed identically for both polymorphs.

The inserted seed is represented by a smooth radial profile,
\begin{equation}
p_s(\mathbf{x})
=
\frac{1}{2}
\left[
1-\tanh
\left(
\frac{|\mathbf{x}-\mathbf{x}_s|-R_s}{w_s}
\right)
\right],
\label{eq:pfm-seed-profile}
\end{equation}
\noindent\ 
where $R_s=15~\text{\AA}$ sets the nominal seed radius and $w_s=5~\text{\AA}$ controls the smoothing length of the diffuse profile. These parameters characterize the prescribed phase-field initialization, whereas $n_{c,i}$, expressed as the number of atoms, denotes the atomistic critical nucleus size used in the nucleation-barrier analysis of Supplementary Note 2.

The profile is inserted into the selected phase field using a maximum operation:
\begin{equation}
\begin{aligned}
\eta_\mathrm{A}(\mathbf{x},t_s^+)
&=
\max
\left[
\eta_\mathrm{A}(\mathbf{x},t_s^-),
p_s(\mathbf{x})
\right],
\qquad
i_s=\mathrm{A},\\
\eta_\mathrm{R}(\mathbf{x},t_s^+)
&=
\max
\left[
\eta_\mathrm{R}(\mathbf{x},t_s^-),
p_s(\mathbf{x})
\right],
\qquad
i_s=\mathrm{R},
\end{aligned}
\label{eq:pfm-seed-max-insertion}
\end{equation}
\noindent\ 
where $t_s^-$ and $t_s^+$ denote the times immediately before and after seed insertion, respectively. The unselected phase field and the composition field $c$ remain unchanged during insertion. The CFM assignment is performed once, after which the phase fields evolve according to the free-energy and evolution equations in Sections 4.3 and 4.4.

\paragraph*{4.3 Phase-field free-energy functional}

The phase-field free-energy functional $F$ is written in terms of the free-energy density $f$ over the simulation domain $\Omega$ as
\begin{equation}
F
=
\int_{\Omega}
f(\eta_\mathrm{A},\eta_\mathrm{R},c,\nabla\eta_\mathrm{A},\nabla\eta_\mathrm{R},\nabla c)
\,d\mathbf{x}.
\label{eq:pfm-total-free-energy}
\end{equation}
\noindent\ 
The free-energy density contains a harmonic composition term, phase-dependent bulk free-energy terms, double-well barriers, an anatase--rutile overlap penalty, and gradient-energy terms:
\begin{equation}
\begin{aligned}
f
=
&\ \frac{1}{2}K_\mathrm{P}(c-c_0)^2
+ h(\eta_\mathrm{A})\Delta f_\mathrm{A}(c)
+ h(\eta_\mathrm{R})\Delta f_\mathrm{R}(c) \\
&+ W_\mathrm{A}g(\eta_\mathrm{A})
+ W_\mathrm{R}g(\eta_\mathrm{R})
+ \chi_{\mathrm{AR}}\eta_\mathrm{A}^{2}\eta_\mathrm{R}^{2} \\
&+ \frac{1}{2}\kappa_\mathrm{A}|\nabla\eta_\mathrm{A}|^2
+ \frac{1}{2}\kappa_\mathrm{R}|\nabla\eta_\mathrm{R}|^2
+ \frac{1}{2}\kappa_c|\nabla c|^2.
\end{aligned}
\label{eq:pfm-free-energy-density}
\end{equation}
\noindent\ 
The interpolation and double-well functions are
\begin{equation}
h(u)=u^2(3-2u),
\qquad
g(u)=u^2(1-u)^2.
\label{eq:pfm-interpolation-double-well}
\end{equation}
\noindent\ 
Here, $h(u)$ smoothly interpolates the phase free-energy contribution between $u=0$ and 1, whereas $g(u)$ penalizes intermediate order-parameter values.

For $i\in\{\mathrm{A},\mathrm{R}\}$, the remaining quantities are:
\begin{itemize}
\item $K_\mathrm{P}$: harmonic composition coefficient; $K_\mathrm{P}=1.0~\mathrm{eV\,\text{\AA}^{-3}}$.
\item $\Delta f_i(c)$: composition-dependent parent-to-polymorph free-energy density from Section 2.2.3 at 1000~K; the nearest boundary value was used outside the evaluated composition range.
\item $W_i$ and $\kappa_i$: double-well and gradient-energy coefficients, respectively, parameterized using the interfacial energies from Section 2.3 and a common diffuse-interface width of $10~\text{\AA}$.
\item $\chi_\mathrm{AR}$: anatase--rutile overlap coefficient; $\chi_\mathrm{AR}=\alpha_\mathrm{AR}\max(W_\mathrm{A},W_\mathrm{R})$ with $\alpha_\mathrm{AR}=5$.
\item $\kappa_c$: composition-gradient coefficient; $\kappa_c=0.5~\mathrm{eV\,\text{\AA}^{-1}}$.
\end{itemize}

\paragraph*{4.4 Evolution equations and numerical implementation}

The anatase and rutile order parameters are nonconserved fields and therefore evolve by Allen--Cahn dynamics:
\begin{equation}
\frac{\partial\eta_\mathrm{A}}{\partial t}
=
-L_\mathrm{A}
\frac{\delta F}{\delta\eta_\mathrm{A}},
\qquad
\frac{\partial\eta_\mathrm{R}}{\partial t}
=
-L_\mathrm{R}
\frac{\delta F}{\delta\eta_\mathrm{R}}.
\label{eq:pfm-allen-cahn}
\end{equation}
\noindent\ 
Here, $L_\mathrm{A}$ and $L_\mathrm{R}$ are the kinetic mobilities of the anatase and rutile phase fields. With constant mobilities, composition affects the phase-field evolution through the local thermodynamic driving terms, the CFM-informed seed identity, and the conserved-field redistribution described below.
For either phase field,
\begin{equation}
\frac{\delta F}{\delta\eta_i}
=
\frac{\partial f}{\partial\eta_i}
-\kappa_i\nabla^2\eta_i,
\qquad
i\in\{\mathrm{A},\mathrm{R}\}.
\label{eq:pfm-eta-variational-derivative}
\end{equation}
\noindent\ 
These equations relax each phase field down the free-energy gradient while preserving separate anatase and rutile growth channels.

The composition field is conserved and evolves by a Cahn--Hilliard equation:
\begin{equation}
\frac{\partial c}{\partial t}
=
\nabla\cdot
\left(
M_c\nabla\mu_c
\right),
\qquad
\mu_c
=
\frac{\delta F}{\delta c}
=
\frac{\partial f}{\partial c}
-\kappa_c\nabla^2c.
\label{eq:pfm-cahn-hilliard}
\end{equation}
\noindent\ 
Here, $M_c$ is the mobility of the conserved composition field and $\mu_c$ is the chemical potential obtained from the same free-energy density. The mobilities were held constant at $L_\mathrm{A}=L_\mathrm{R}=5.0$ and $M_c=0.25$ in model units; their common scale defines the model time unit.

The phase-field equations were solved for 2000 time steps with $\Delta t=0.01$ using a semi-implicit Euler scheme, with spatial derivatives evaluated in Fourier space under periodic boundary conditions. Following seed insertion and after each time step, the phase fields were projected to satisfy $\eta_\mathrm{A}\ge 0$, $\eta_\mathrm{R}\ge 0$, and $q=\eta_\mathrm{A}+\eta_\mathrm{R}\le 1$, so that the parent fraction $\phi_\mathrm{P}=1-q$ remains nonnegative. The composition field was shifted uniformly after each time step to maintain its prescribed spatial mean.


\clearpage

\begin{figure}
  \centering
  \includegraphics[width=0.9\textwidth]{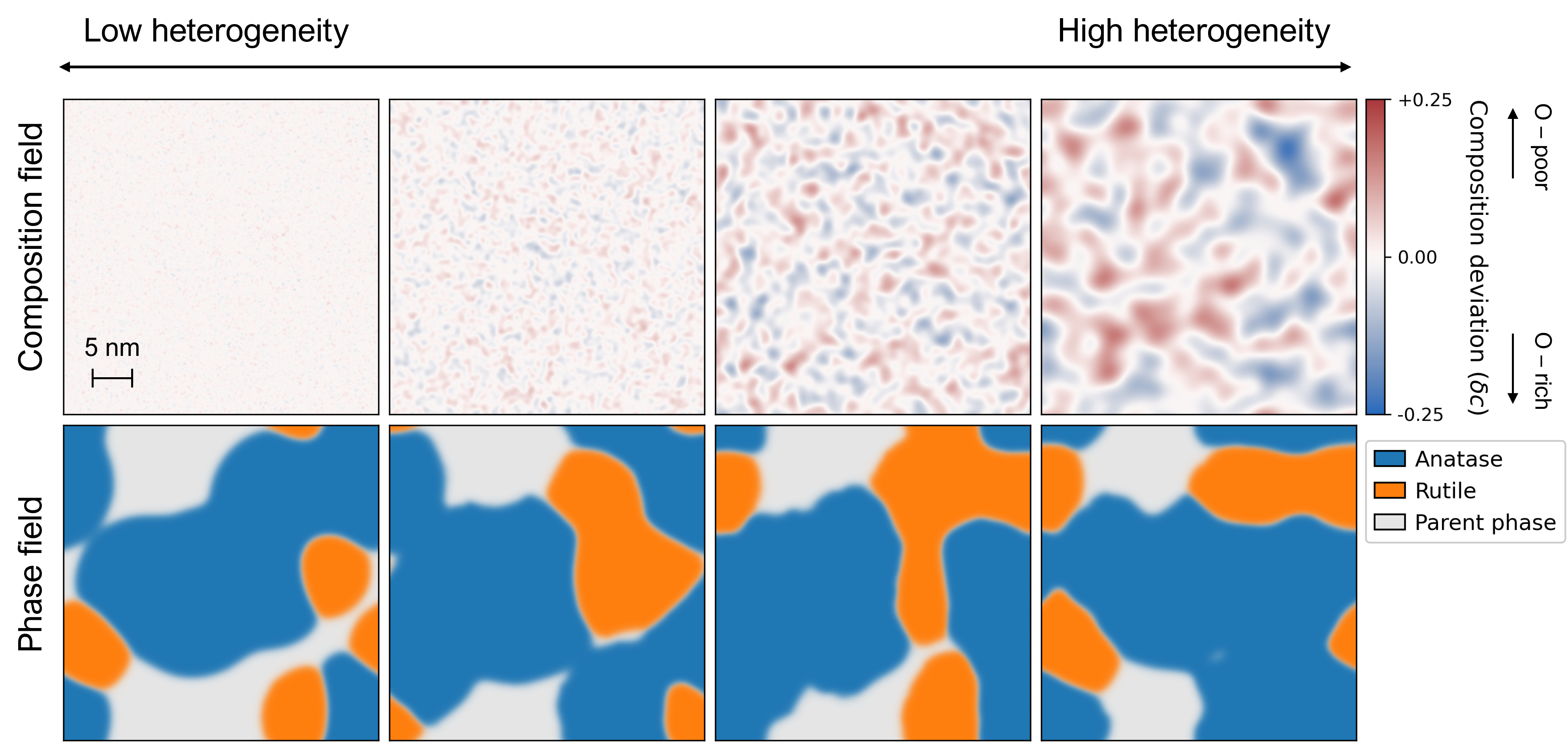}
  \caption{\small\textbf{Phase-field simulation from heterogeneous composition fields for the TiO$_{1.98}$ case.} The top and bottom rows show the initial composition fields and corresponding phase fields after evolution, respectively, with heterogeneity increasing from left to right. The color scale shows the local deviation in normalized oxygen composition from its spatial mean ($\delta c$); positive (red) and negative (blue) values correspond to O-poor and O-rich local environments, respectively. Blue, orange, and light gray denote anatase, rutile, and the parent phase, respectively. The 5 nm scale bar applies to all panels.}
  \label{fig:supp-phase-field-tio1p98}
\end{figure}

\begin{figure}
  \centering
  \includegraphics[width=0.9\textwidth]{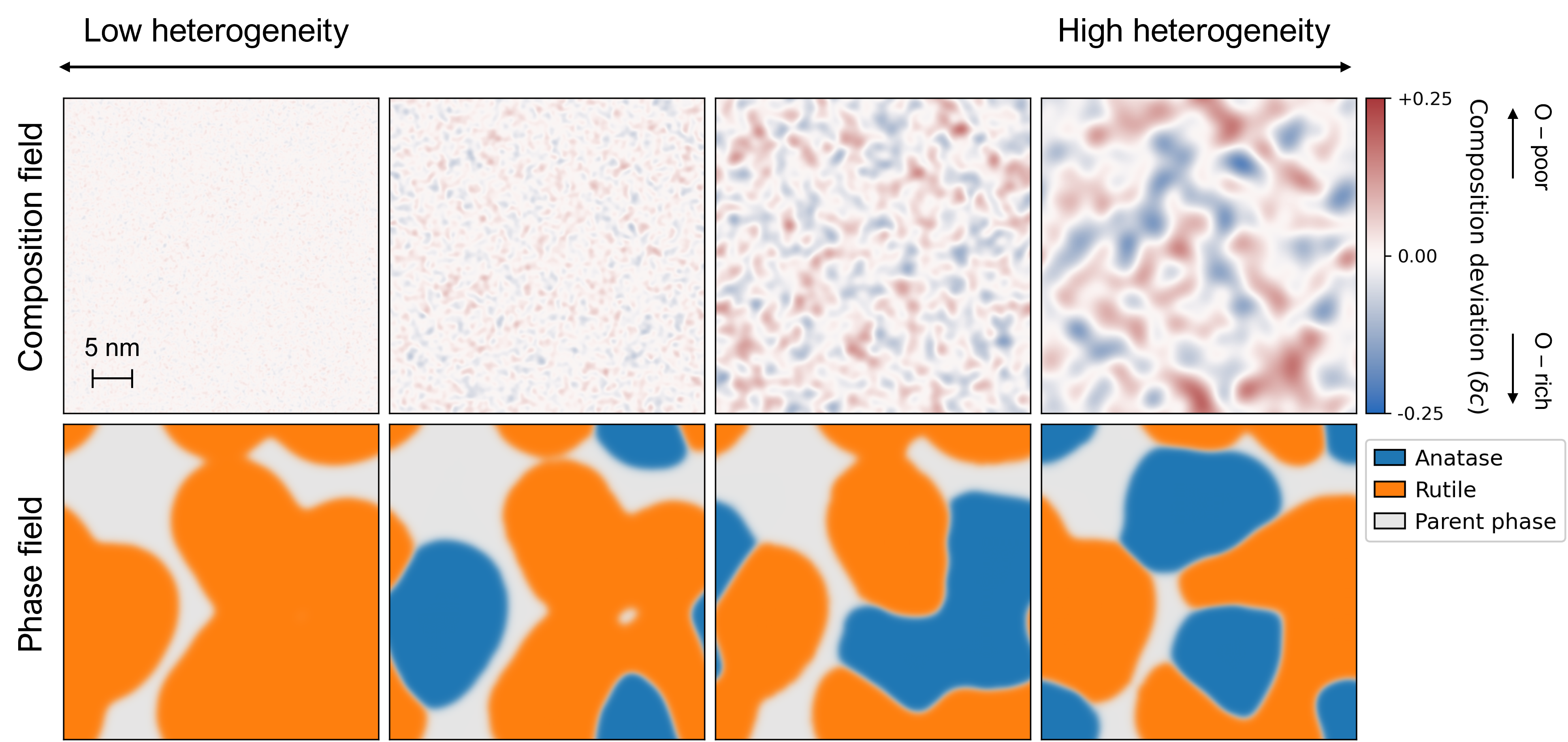}
  \caption{\small\textbf{Phase-field simulation from heterogeneous composition fields for the TiO$_{1.96}$ case.} The top and bottom rows show the initial composition fields and corresponding phase fields after evolution, respectively, with heterogeneity increasing from left to right. The color scale shows the local deviation in normalized oxygen composition from its spatial mean ($\delta c$); positive (red) and negative (blue) values correspond to O-poor and O-rich local environments, respectively. Blue, orange, and light gray denote anatase, rutile, and the parent phase, respectively. The 5 nm scale bar applies to all panels.}
  \label{fig:supp-phase-field-tio1p96}
\end{figure}

\begin{figure}
  \centering
  \includegraphics[width=0.9\textwidth]{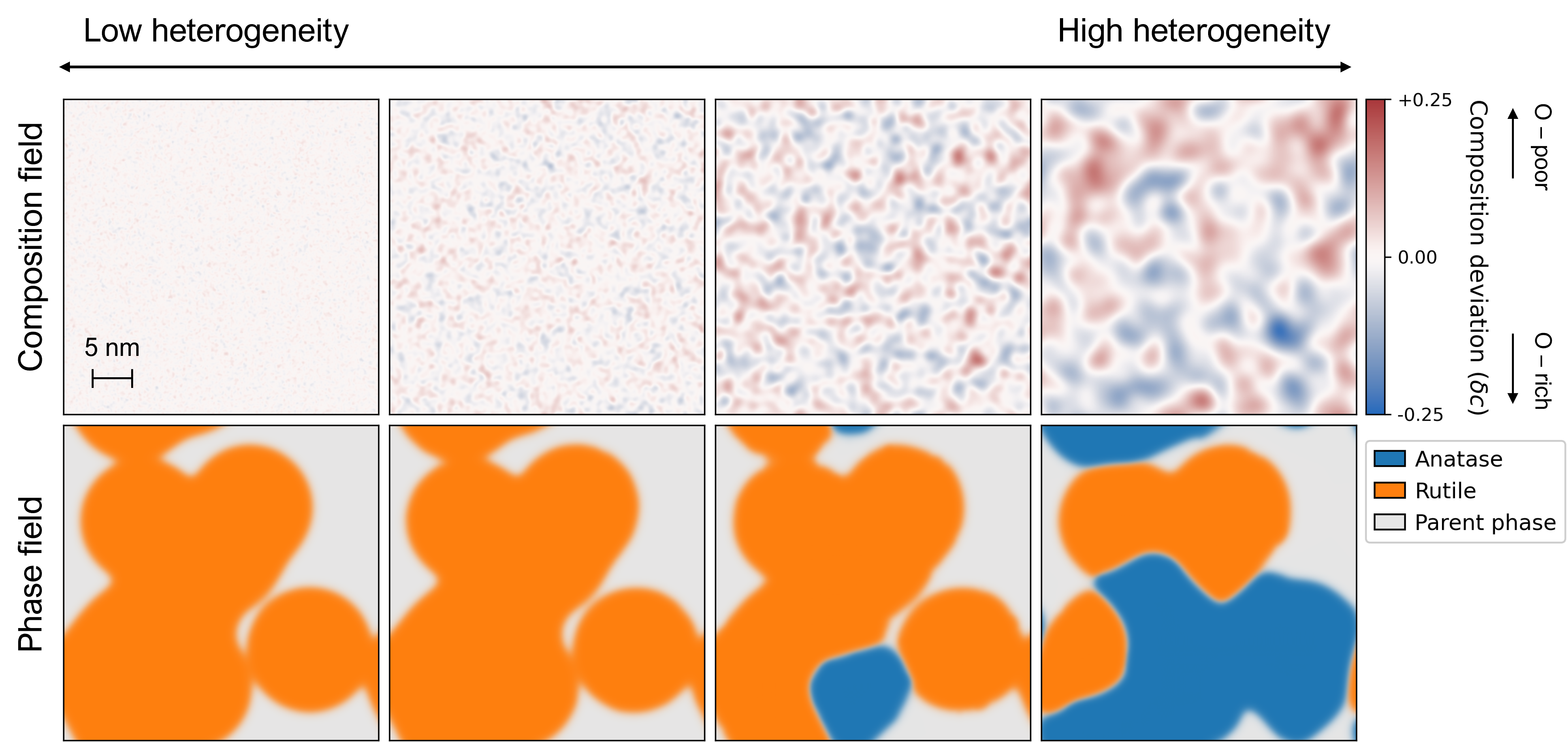}
  \caption{\small\textbf{Phase-field simulation from heterogeneous composition fields for the TiO$_{1.94}$ case.} The top and bottom rows show the initial composition fields and corresponding phase fields after evolution, respectively, with heterogeneity increasing from left to right. The color scale shows the local deviation in normalized oxygen composition from its spatial mean ($\delta c$); positive (red) and negative (blue) values correspond to O-poor and O-rich local environments, respectively. Blue, orange, and light gray denote anatase, rutile, and the parent phase, respectively. The 5 nm scale bar applies to all panels.}
  \label{fig:supp-phase-field-tio1p94}
\end{figure}


\clearpage

\paragraph{Caption for Data S1.}
\textbf{Literature-reported anatase--rutile phase observations used in Fig. 2C.}
Data S1 contains the compiled experimental observations plotted in Fig. 2C, including the indexed source, temperature, pressure, reported phase, and source metadata.

\paragraph{Caption for Data S2.}
\textbf{Experimental scatter data used in Fig. 5.}
Data S2 contains the experimental scatter points plotted in Fig. 5, including the indexed source, temperature, oxygen partial pressure, reported phase, and source metadata.

\ifdefined\CombinedSubmission
\else

\bibliographystyle{sciencemag}
\bibliography{references}
\fi

\FinishSupplement

\end{document}